\documentclass[amsmath,amssymb,amsbsy,twocolumn,aps,prb,reprint,superscriptaddress,showpacs]{revtex4-1}

\usepackage{amsfonts} 
\usepackage{amsmath}
\usepackage{amssymb}
\usepackage{cancel} 
\usepackage{bm}
\usepackage[usenames, dvipsnames]{color} 
\usepackage{dcolumn}
\usepackage{epic} 
\usepackage{epsfig}
\usepackage{eucal} 
\usepackage{feynmp}    
\usepackage{graphicx}
\usepackage[breaklinks,colorlinks = true,linkcolor = red,urlcolor=cyan,citecolor=red]{hyperref}
\usepackage{mathrsfs}
\usepackage{microtype}
\usepackage{soul}
\usepackage{subfigure}
\usepackage{wrapfig} 
\usepackage{wasysym}
\usepackage{xy} 
\usepackage{natbib}

\newcommand{\vect}[1]{\boldsymbol{#1}}

\begin{document}


\title{Electric field control of emergent electrodynamics in quantum spin ice}

\author{\'{E}tienne Lantagne-Hurtubise}
\email{lantagne@phas.ubc.ca}
\affiliation{Max-Planck-Institut f{\"u}r Physik komplexer Systeme, N{\"o}thnitzer Str. 38, 01187 Dresden, Germany}
\affiliation{Department of Physics and Astronomy, University of British Columbia, Vancouver,  BC, V6T 1Z1, Canada}

\author{Subhro Bhattacharjee}
\affiliation{International Centre for Theoretical Sciences, Tata Institute of Fundamental Research, Bangalore 560012, India}
\author{R. Moessner}
\affiliation{Max-Planck-Institut f{\"u}r Physik komplexer Systeme, N{\"o}thnitzer Str. 38, 01187 Dresden, Germany}

\begin{abstract}

We study the coupling between conventional (Maxwell) and emergent electrodynamics in quantum spin ice, a 3+1-dimensional $U(1)$ quantum spin liquid.  We find that a uniform electric field can be used to tune the properties of both the ground state and excitations of the spin liquid. In particular, it induces \textit{emergent birefringence}, rendering the speed of the emergent light anisotropic and polarization-dependent. A sufficiently strong electric field triggers a quantum phase transition into new $U(1)$ quantum spin liquid phases which trap emergent electric $\pi$-fluxes. The flux patterns of these new phases depend on the direction of the electric field. Strikingly, some of the canonical \textit{pinch points} in the spin structure factor, characteristic of classical spin ice, emerge near the phase transition, while they are absent in the quantum spin liquid phases. Estimating the electric field strength required, we find that this transition is potentially accessible experimentally.  Finally, we propose a minimal mechanism by which an \textit{oscillating} electric field can generate emergent radiation inside a quantum spin ice material with non-Kramers spin doublets.

\end{abstract}
\date{\today}
\maketitle
\section{Introduction}
Understanding the intertwined electric and magnetic properties of a solid is of great importance, not only for technology (such as in the context of  multiferroicity\cite{Spaldin391}), but also from a fundamental physics viewpoint. Here, we address the question of how a quantum spin liquid (QSL) described by an emergent $U(1)$ gauge field -- essentially, an emergent form of Maxwell electromagnetism -- responds to the application of a laboratory {\em electric} field. The underlying question is thus how the fundamental electromagnetism of the vacuum interacts with the one generated dynamically by the magnetic degrees of freedom giving rise to the QSL. 

 Our model system is quantum spin ice\cite{PhysRevB.69.064404,gingras2014quantum,PhysRevB.68.054405,PhysRevLett.91.167004,savary2012coulombic,castelnovo2008magnetic,PhysRevB.86.075154,PhysRevLett.108.067204,henley2010coulomb,lee2012generic,PhysRevLett.80.2933,PhysRevLett.108.247210,PhysRevB.68.184512} -- a three dimensional quantum spin liquid.\cite{wen2002quantum,kitaev2006anyons,PhysRevLett.86.1881,lee2008physics} Owing to our growing understanding of such {\it quantum ordered}\cite{wen2004quantum} phases and, in parallel,  the discovery of a number of frustrated rare-earth pyrochlore magnets that do not order to the lowest experimentally accessible temperatures, \cite{pan2015measure,lhotel2014first,chang2012higgs,applegate2012vindication,thompson2011rods,ross2011quantum,yin2013low,fennell2012power,legl2012vibrating,pan2015measure}  the study of  quantum spin ice  has received much impetus as a rare condensed matter physics  example of a three-dimensional state beyond the conventional Landau-Ginzburg-Wilson description of the phases of matter. 

The $U(1)$ gauge structure\cite{PhysRevB.68.184512,PhysRevB.69.064404,PhysRevLett.91.167004}  of quantum spin ice manifests itself  as an emergent electromagnetism comprising gapless collective emergent `photons', alongside novel gapped excitations -- emergent `electric charges' and  `magnetic monopoles'. [In the following, we will drop the inverted commas around the emergent analogues of electromagnetic entities.] In particular, the gaplessness of the photons, resulting from the emergent $U(1)$ gauge structure,  is expected to be more robust than  that of Nambu-Goldstone bosons arising from the spontaneous  breaking of a continuous symmetry\cite{Goldstone1961}: \textit{e.g.}, magnons in a magnetically ordered state  are generically gapped out  by perturbations breaking the continuous symmetry down to a discrete one.  
  
How, then, can one probe and manipulate these new types of excitations? As far as the magnetic monopoles are concerned,  it has been proposed to model their response to a magnetic field along the same lines as the response of an electrolyte to an  electric field.\cite{castelnovo2008magnetic,bramwell2009measurement}  Via such a magnetolyte/magnetricity analogy, several features -- such as a magnetic  liquid-gas transition,\cite{castelnovo2008magnetic} and even the nonlinear AC susceptibility \cite{kaiser2013onsager,ryzhkin}-- can be naturally modeled in classical spin ice. 

In this paper, we address the magnetoelectric properties of quantum spin ice. By considering the effect of an external electric field, we show that both the ground state and excitations respond in a characteristic fashion.  In particular, we show that -- (1)  the emergent photons acquire a \textit{birefringent behavior} which depends sensitively on the external electric field direction, and (2) sufficiently strong fields can change the nature of the quantum spin ice ground state by stabilizing different types of $U(1)$  QSLs with non-trivial distributions of the emergent electric flux.  These findings are related in that the emergent speed of light vanishes  for certain combinations of the applied electric field and the polarization/propagation direction of the emergent photons, rendering the zero-flux QSL unstable. This triggers a (presumably discontinuous) quantum phase transition to an  anisotropic $\pi$-electric flux $U(1)$ QSL, where the flux distribution is related to the direction of the applied electric field. 
 
This provides a new perspective on the question of how to identify the presence of  a $U(1)$ QSL  in candidate rare-earth pyrochlores, particularly on how to find concrete signatures of gapless emergent photons and manipulate them through external probes. It has already been noted that the spin structure factors probed in neutron scattering experiments exhibit features characteristic to U(1) QSLs that are due to the emergent photons\cite{castelnovo2012spin, PhysRevB.69.064404, PhysRevB.86.075154}. However, these features have a vanishing intensity at low energies. Such vanishing of intensity in the equal-time spin structure factor of quantum spin ice should be contrasted with the finite intensity characteristic {\it pinch points} in the static spin structure factor of classical spin ice.\cite{castelnovo2012spin, PhysRevB.23.232, PhysRevLett.93.167204, PhysRevB.71.014424}  In this work, we show that a subset of the canonical pinch points re-emerge as the `photon' velocity vanishes near the phase transition mentioned above. Crucially, we estimate that this transition could potentially be observed with experimentally accessible electric fields and for typical values of the coupling constants in candidate quantum spin ice materials.  In addition, we  point out an interesting  surface effect in quantum spin ice made of non-Kramers spin doublets, where shining an oscillating electric field on the surface can generate radiation of emergent light inside the material. 
 
Our analysis of the magnetoelectric effects in quantum spin ice is based on the theory of the coupling of electric fields  with spins in magnetic insulators. Due to virtual charge fluctuation and/or magnetostriction, magnetic insulators can develop electric  polarization which is a function of the spin degrees of freedom.\cite{PhysRevB.78.024402} This electric polarization in turn couples to electric fields. Since the nature of the coupling to an electric field is very different from the standard Zeeman-type coupling  to a magnetic field, it may produce  qualitatively different low-energy responses such as charge response ({\it e.g.}, optical conductivity\cite{PhysRevB.78.024402,hwang2014signatures}). This can be a particularly interesting way to understand the low-energy behavior in frustrated magnets with possible QSL ground states and unconventional excitations.\cite{PhysRevB.87.245106,PhysRevB.87.235108} Experimentally, an interesting power-law dependence of optical conductivity on frequency has been observed in the kagome QSL candidate Herbertsmithite.\cite{PhysRevLett.111.127401} 

In the context of classical spin ice systems, Khomskii's seminal work\cite{khomskii2012electric} found that the gapped emergent magnetic monopoles carry an electric dipole moment which influences the dielectric response of the material \cite{PhysRevB.72.144422} and paves the way for the appearance of monopole crystals\cite{PhysRevB.91.214422}. However, this leaves open the question of the effect of electric fields below the monopole excitation energy scale. Such effects are absent in classical spin ice,\cite{khomskii2012electric} but, as we show here, can lead to non-trivial features in quantum spin ice. In this vein, this present work extends Khomskii's\cite{khomskii2012electric}  by taking into account  (a) the symmetry of the rare-earth pyrochlores, and (b) quantum effects which lead to qualitatively novel behavior below the monopole excitation energy scale.

\begin{figure*}
	\centering
	\subfigure[]{
		\includegraphics[width=0.82\textwidth]{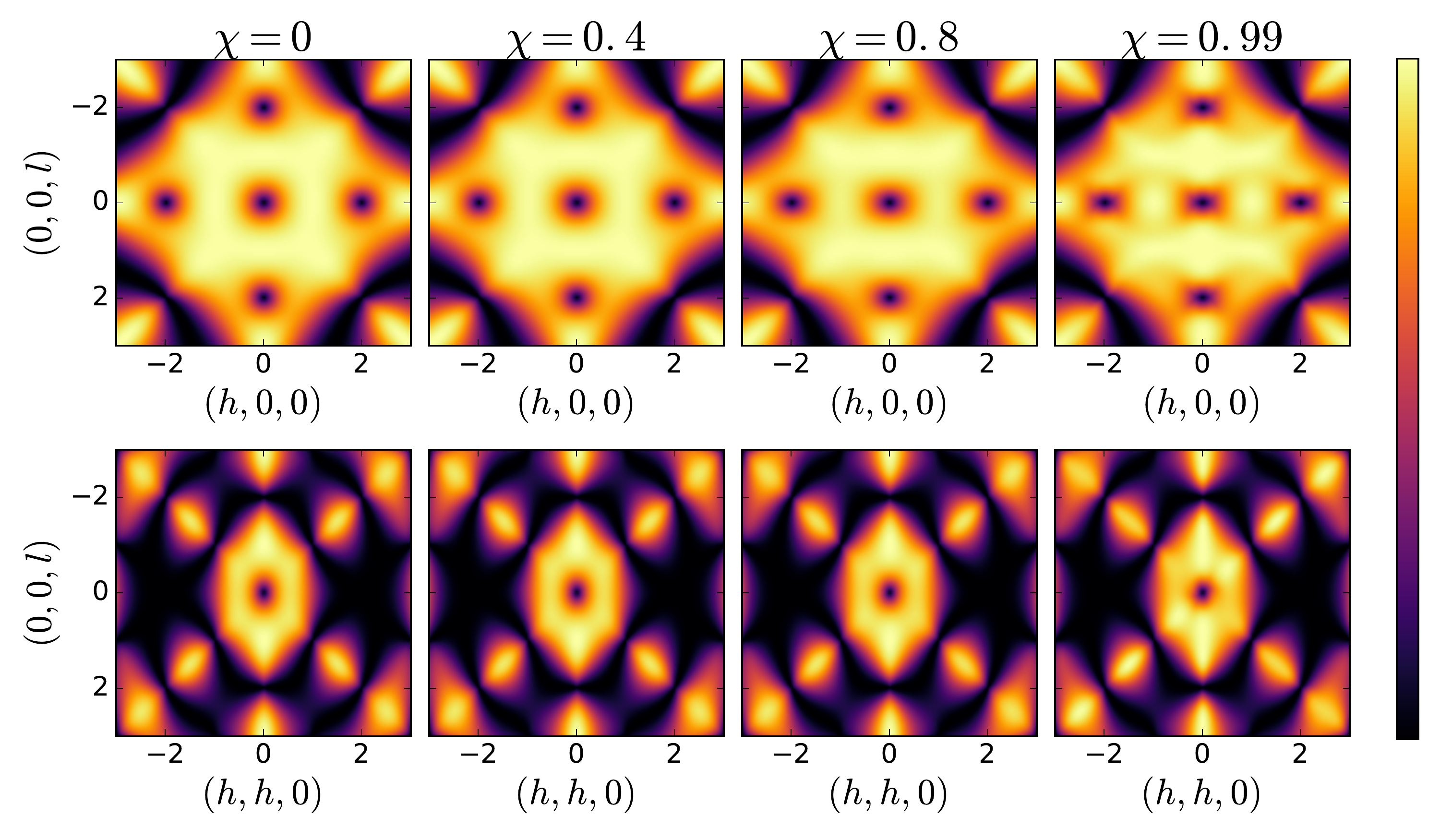}
		\label{fig_structurefactors_1}
	}	
	\subfigure[]{
		\includegraphics[width=0.82\textwidth]{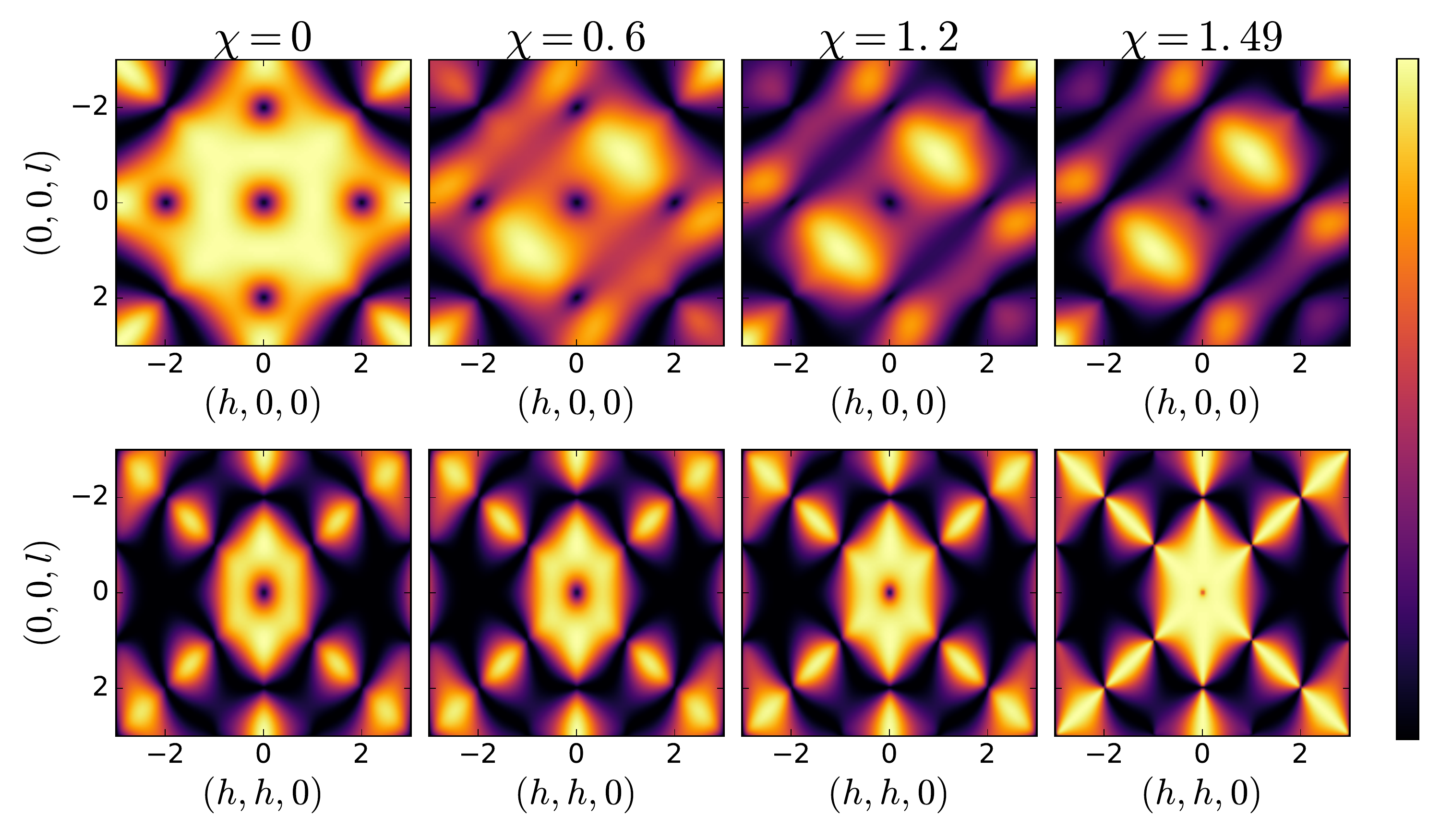}
		\label{fig_structurefactors_2}
	}
	\caption{ Equal-time spin structure factors in the spin-flip channel (obtained from Eq. (\ref{eq_structurefactors})), for (a): ${\bf E} \propto [011]$ and (b): ${\bf E} \propto [111]$ respectively. The top panels show the $[h0l]$ scattering plane (using polarized neutrons with $\vect{\eta}_\nu \propto [010]$), and the bottom panels show the $[hhl]$ scattering plane (using polarized neutrons with $\vect{\eta}_\nu \propto [1\overline{1}0]$). We use $\mathcal{W}/2Ug= 0.1$ to regularize the theory near the phase transition, as explained in Sec. \ref{sec_transition} (see Eqs. (\ref{eq_dispersion_reg}) and (\ref{eq_dispersion_reg_2})). Each subplot has an independent color scale (see Sec. \ref{sec_transition} for a discussion of the intensity of the structure factors). }
	\label{fig_structurefactors}
\end{figure*}

The remainder of this paper is organized as follows. {For completeness, we begin with a short description of the minimal spin model for quantum spin ice in Sec. \ref{sec_quantum spin iceham} and discuss the limit in which it supports a $U(1)$ QSL ground state. In Sec. \ref{sec_pol} we introduce the electric polarization operator and derive its explicit form using symmetries. We then obtain the low-energy effective theory of quantum spin ice in the presence of a uniform external electric field. We study the properties of this theory for small electric fields in Sec. \ref{sec_biref}.  We find that the speed of the emergent photons becomes both polarization- and direction-dependent. This electric field-induced birefringence leads to experimentally observable  changes in the spin structure factor, as shown in Fig. \ref{fig_structurefactors}. Sec. \ref{sec_symmetrybreaking} describes the trapping of (emergent) electric  $\pi$-fluxes for large external electric fields.  The QSLs with and without $\pi$-fluxes (at high and low external electric fields, respectively) represent different $U(1)$ QSLs where the ground state and the low-energy excitations transform under different projective representations of the symmetry group. Hence, they are separated by a quantum phase transition, presumably first order, as we argue in Sec. \ref{sec_transition}. In Sec. \ref{sec_materials}, we briefly discuss the relevance of our work to candidate quantum spin ice materials and estimate the electric field strengths needed to observe the effects mentioned above. Finally, we comment on a potentially interesting surface effect of an oscillating external electric field on quantum spin ice materials with non-Kramers doublets in Sec. \ref{sec_surface}. We conclude with a brief summary of the present work in Sec. \ref{sec_summary}. The details of various calculations are discussed in the appendices. 

\section{The Quantum spin Ice Hamiltonian}
\label{sec_quantum spin iceham}

In candidate quantum spin ice materials, typically rare-earth pyrochlores with the chemical formula $R_2T_2$O$_7$, the rare earth (R) magnetic moments sit on a network of corner-sharing tetrahedra (see Appendix \ref{appen_pyrochlorelattice}).  Due to the  interplay of spin-orbit coupling, crystal field and Coulomb repulsion between electrons, the low-energy magnetic degree of freedom, often an Ising doublet, has natural quantization axes changing from site to site and pointing along the local $[111]$ crystallographic direction.\cite{gingras2014quantum}  Choosing, without loss of generality, the up tetrahedra to define the axes of spin quantization, the spins can be denoted as ${\bf S}_i=\hat{\bf t}_i s^z_i+\hat{\bf x}_i s^x_i+\hat{\bf y}_i s^y_i$, where $\hat{\bf t}_i, \hat{\bf x}_i$ and $\hat{\bf y}_i$ form local triads\cite{savary2012coulombic} (see Appendix \ref{appen_latticedetails}). The minimal Hamiltonian, consistent with symmetries, that may stabilize the quantum spin ice is given by
\begin{align}
H=&J_{zz}\sum_{\left< ij\right>} s^z_is^z_j-J_\pm\sum_{\left< ij\right>}(s^+_is^-_j+s^-_is^+_j),
\label{eq_quantum spin iceham}
\end{align}
where $J_{zz}, J_\pm >0$. In the regime where $J_{zz}\gg J_\pm$, the ground state of Eq. (\ref{eq_quantum spin iceham}) is a $U(1)$ QSL -- the quantum spin ice, with gapped electric and magnetic charges and gapless photons (see below).\cite{PhysRevB.69.064404} While other terms consistent with symmetries are allowed and may be of importance to actual rare-earth materials,\cite{savary2012coulombic} to simplify calculations and explore the magnetoelectric effects in the quantum spin ice state, we focus on Eq. (\ref{eq_quantum spin iceham}) in the rest of this paper.

The low-energy description of the QSL phase can be obtained by starting with the Hamiltonian in Eq. (\ref{eq_quantum spin iceham})  and performing a perturbative expansion in $J_\pm/J_{zz}$ to the leading non-trivial orders. While this is well known,\cite{PhysRevB.69.064404, gingras2014quantum, PhysRevB.68.184512} we briefly review it here for the sake of completeness and also setting up our notations.
\subsection{Low-energy theory of quantum spin ice}

The classical ground states favored by the $J_{zz}$ term consist of two spins pointing in and two spins pointing out of every tetrahedron. This two-in-two-out ice manifold is macroscopically degenerate and is separated from the excited states (in which at least two tetrahedra do not satisfy the ice rules) by an energy of the order of $J_{zz}$.\cite{gingras2014quantum} The transverse terms lift this degeneracy without leading to magnetic ordering, resulting in the quantum spin ice state.\cite{PhysRevB.68.184512, PhysRevB.69.064404}

The leading-order non-trivial term in the degenerate perturbation theory, as shown by Hermele {\it et al.}\cite{PhysRevB.69.064404} (also see Appendix \ref{appen_pert1}), comes from cooperative flipping of the spins along the smallest closed loops on the pyrochlore lattice: the hexagons formed by six tetrahedra. This leads to an effective low-energy Hamiltonian given by 
\begin{align}
\mathcal{H}_{\text{eff}}=&-g\sum_{{\hexagon}}\left(\mathcal{O}_{\hexagon}+\text{h.c.}\right),
\label{eq_eff_ham0}
\end{align}
where $\mathcal{O}_{\hexagon}=s_1^+s_2^-s_3^+s_4^-s_5^+s_6^-$ ($1,\dots 6\in {\hexagon}$) is an operator that flips six spins on a given hexagonal loop,  and $g = 12 J_\pm^3/J_{zz}^2$.   

 The low-energy physics of the system encoded within the effective Hamiltonian [Eq. (\ref{eq_eff_ham0})] becomes transparent following the mapping to an effective problem of electromagnetism.\cite{PhysRevB.69.064404} To this end, we note that each site of the pyrochlore lattice can be uniquely identified with a bond of the medial diamond lattice that is obtained by joining the centers of the tetrahedra forming the pyrochlore lattice (see Appendix \ref{appen_pyrochlorelattice}). The spins sit on the bonds of this diamond lattice. For a spin at site $i$ of the pyrochlore lattice, we write
\begin{align}
s^z_i=b_{\bf rr'}-\frac{1}{2}, \quad s_i^{\pm}=e^{\pm i \alpha_{\bf rr'}},
\end{align}
where ${\bf r}$ (${\bf r'}$) denotes the center of an up (down) tetrahedron, $b_{\bf rr'}$ is the emergent magnetic field, and $\alpha_{\bf rr'}$ is the dual vector potential that is conjugate to $b_{\bf rr'}$, {\it i.e.} on a lattice :
\begin{align}
[b_{\bf rr'},\alpha_{\bf r''r'''}]=i\left(\delta_{\bf rr''}\delta_{\bf r'r'''}-\delta_{\bf rr'''}\delta_{\bf r'r''}\right).
\end{align}
The emergent electric field, 
\begin{align}
e_{\bf ss'}= \nabla_{\hexagon} \times  \alpha_{\bf rr'},
\end{align}
is given by the lattice curl of the dual vector potential around the hexagonal loops of the pyrochlore lattice. $e_{\bf ss'}$, therefore, is defined on the links of the dual diamond lattice which are denoted by ${\bf ss'}$ in accordance with the right-hand rule \footnote{Note that our definitions of the emergent fields $e$ and $b$ follow Ref. \onlinecite{PhysRevB.86.075154}, and are reversed with respect to Ref. \onlinecite{PhysRevB.69.064404}.}.

In terms of these new variables,  the effective low-energy Hamiltonian [Eq. (\ref{eq_eff_ham0})] can be recast as\cite{PhysRevB.69.064404}
\begin{align}
\mathcal{H}_{\text{eff}}=&\frac{U}{2}\sum_{\left< \bf rr'\right>}{ b}_{\bf rr'}^2 - 2g \sum_{\left<\bf ss'\right>} \cos[{ e}_{\bf ss'}].
\label{eq_lattice_ham0}
\end{align}
with $U>0$ a model parameter. This low-energy effective theory is a pure compact $U(1)$ lattice gauge theory in (3+1) dimensional space-time. The deconfined phase of this theory corresponds to a $U(1)$ quantum spin liquid phase, which is of interest to us. In this deconfined phase, the compactness of the $U(1)$ gauge group is not important and we can expand the cosine terms to get a more explicit similarity with the theory of quantum electrodynamics\cite{PhysRevB.69.064404}: up to a constant, 
	\begin{align}
		\mathcal{H}_{\text{eff}}=&\frac{U}{2}\sum_{\left< \bf rr'\right>}{ b}_{\bf rr'}^2 + g \sum_{\left<\bf ss'\right>} { e}_{\bf ss'}^2.
		\label{eq_lattice_ham0_expanded}
	\end{align}
While we can continue working with Eq. (\ref{eq_lattice_ham0_expanded}) (as we show below), it is insightful to first derive a continuum limit of this theory by taking
\begin{align*}
\sum\rightarrow \frac{1}{l^3}\int d^3{\bf r}, \quad e_{\mathbf{ss'}} = l {\bf e\cdot \hat l_{ss'}}, \quad b_{\bf rr'}=l{\bf b\cdot \hat t_{rr'}},
\end{align*}
where $l$ is a lattice length scale (note that this is a slightly different way of scaling than used in Ref. \onlinecite{PhysRevB.69.064404}). Here, ${\bf{\hat l}_{ss'}}$ is the unit vector in the direction from $\mathbf{s}$ to $\mathbf{s'}$ on the dual diamond lattice, which is in one-to-one correspondence with the quantization axes of spin ice denoted by $\hat{\mathbf{t}}_m$. More precisely,
$$ e_{\mathbf{ss'}} = l {\bf e\cdot \hat{t}_m} \quad \text{for} \quad \left< {\bf ss'}\right>\parallel \hat{\bf t}_m .$$
We stress that the spin ice quantization axes $ \hat{\mathbf{t}}_m $ do \textit{not} form an orthonormal basis, but instead respect the condition $\sum_m \hat{t}_m^\alpha \hat{t}_m^\beta  =  \frac{4}{3} \delta^{\alpha \beta}$ (see Appendix \ref{appen_latticedetails}).  Using this, the continuum Hamiltonian is  given by: 
\begin{align}
\mathcal{H}_{\text{eff}}^{\rm continuum}=\frac{1}{2}\int d^3{\bf r}\left[\mathcal{U}~{\bf b}^2+\mathcal{K}_0 {\bf e}^2\right]
\label{eq_cont_ham0},
\end{align}
with $\mathcal{U} = 4U/3l$ and $\mathcal{K}_0 = 8g/3l$. This is now the continuum Hamiltonian for a noncompact $U(1)$ gauge theory, similar to the theory of quantum electromagnetism, which supports gapless excitations akin to photons. In the following, we refer to such excitations as emergent photons.

This completes our discussion of the low-energy description of quantum spin ice. We next formulate the problem of applying an external electric field to such a quantum spin ice state, starting with a description of the electric polarization operator.

\section{Electric polarization in quantum spin ice}
\label{sec_pol}

It was recently shown by Bulaevskii {\it et al.} \cite{PhysRevB.78.024402} that virtual charge fluctuations and/or magnetostriction can lead to electric polarization in Mott insulators. Such polarization is a function of the spin operators with the appropriate symmetry, {\it i.e.} a polar vector under lattice symmetries and even under time reversal. Several recent works have studied electric polarization allowed by lattice symmetries, investigating optical conductivity in QSLs\cite{PhysRevB.87.245106,PhysRevB.87.235108} as well as dimerized phases\cite{hwang2014signatures}, leading to interesting predictions for electric field responses.

This problem was recently studied in the context of classical pyrochlore magnets by Khomskii\cite{khomskii2012electric}, who pointed out that the magnetic monopoles of classical spin ice\cite{castelnovo2008magnetic} carry electric dipole moments as they break inversion symmetry. Thus magnetic monopoles couple to external electric fields, affecting dielectric properties of classical spin ice at finite temperature. In view of investigating the magnetoelectric properties of \textit{quantum} spin ice, we note two limitations of this calculation -- (1) the form of the electric polarization operator derived in Ref. \onlinecite{khomskii2012electric} is valid only in the presence of spin-rotation symmetry, which is generally not present in rare-earth pyrochlores (a promising class of candidate quantum spin ice materials), and (2) within the ice manifold, the polarization operator for classical spin ice is identically zero. Hence, there is no polarization effect below the magnetic monopole excitation energy scale.

Here we address these issues explicitly by taking into account both the true microscopic symmetries of rare-earth pyrochlores and quantum effects. We show that these two issues are intricately related to the non-trivial magnetoelectric effects in quantum spin ice below the magnetic monopole energy scale.

The electric polarization operator can in principle be derived, order by order, using strong coupling perturbation theories taking into account the microscopic mechanisms inducing the polarization such as virtual charge fluctuations and/or magnetostriction. However, symmetry considerations alone fix the form of such operators, whereas the bare magnitude of the coupling constants depends on the underlying microscopic mechanisms.\cite{hwang2014signatures,PhysRevB.78.024402} 


\begin{figure}
	\centering
	\includegraphics[width=0.72\columnwidth]{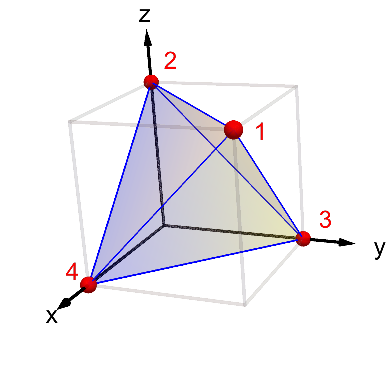}
	\caption{A single up tetrahedron with the global coordinate axes. The components of the electric polarization operator for a single tetrahedron are given by Eq. (\ref{eq_polop}). }
	\label{fig_pyro}
\end{figure}

\subsection{Electric polarization operator for cubic pyrochlores with spin-orbit coupling}

For cubic pyrochlores, the symmetry group is the octahedral group, $O_h\simeq \mathcal{T}_d\bigotimes I$, where $\mathcal{T}_d$ is the tetrahedral group and $I$ is inversion. The irreducible representations admit a triplet $T_{1u}$ which is odd under inversion. Any operator transforming as a $T_{1u}$ representation and even under time reversal is then proportional to the electric polarization operator. The magnetostriction mode that can generate such polarization consists of a staggered trigonal distortion of the up and the down tetrahedra. As noted above, due to the absence of spin-rotation symmetry in the underlying microscopic Hamiltonian of quantum spin ice materials ({\it e.g.}, Eq. (\ref{eq_quantum spin iceham})), the resulting polarization operator is not spin-rotation invariant and hence differs from the form  obtained in Ref. \onlinecite{khomskii2012electric}. The details of the symmetry transformations are given in Appendix \ref{appen_polar}. 

For a single up tetrahedron (shown in Fig. \ref{fig_pyro}), we find it convenient to split the polarization operator  into -- (a) a `classical' or longitudinal (L) contribution, and (b) a `quantum' or transverse (T) contribution. The net polarization operator is given by
\begin{align}
{\bf P}={\bf P}^{(L)}+{\bf P}^{(T)},
\label{eq_polop}
\end{align}
where the three classical/longitudinal components of the electric polarization vector are given by
\begin{align}
{P}^{(L)}_x &= A (s_1^zs_4^z-s_2^zs_3^z)\nonumber,\\
{P}^{(L)}_y &= A(s_1^zs_3^z-s_2^zs_4^z)\nonumber,\\
{P}^{(L)}_z &= A (s_1^zs_2^z-s_3^zs_4^z),
\label{eq_plong}
\end{align}
where $A$ is a coupling constant that depends on the underlying microscopics. Note that these are purely classical terms which are identically zero in the classically degenerate ground-state (ice) manifold. Thus, in order to have non-trivial electric field effects in classical spin ice, one has to incorporate magnetic monopoles, which are high-energy excitations \cite{khomskii2012electric}.
 
In quantum spin ice, however, 
in addition to the above classical terms, 
we find contributions to the polarization operator from the transverse components of the spins. These quantum/transverse contributions are given by
\begin{widetext} 
\begin{align}
{P}_x^{(T)}=&L(-s_1^x-s_4^x+s^x_2+s^x_3) + B\left[(s^+_1s^-_4+\text{h.c.})-(s^+_2s^-_3+\text{h.c.})\right]\nonumber\\
&+ C\left[(\omega s^+_1s^+_4+\text{h.c.})-(\omega s^+_2s^+_3+\text{h.c.})\right]
+D\left[\{s^z_1(\omega^2s^+_4+\omega s_4^-)+(1\leftrightarrow 4)\}-\{s^z_2(\omega^2s^+_3+\omega s^-_3)+(2\leftrightarrow 3)\}\right],
\label{eq_ptrans1}
\end{align}
\begin{align}
{P}_y^{(T)}=&L\left[\frac{1}{2}(s^x_1+s^x_3-s^x_2-s^x_4)-\frac{\sqrt{3}}{2}(s_1^y+s_3^y-s_2^y-s_4^y)\right] 
+B\left[(s^+_1s^-_3+\text{h.c.})-(s^+_2s^-_4+\text{h.c.})\right]\nonumber\\
&+C\left[(\omega^2 s^+_1s^+_3+\text{h.c.})-(\omega^2 s^+_2s^+_4+\text{h.c.})\right]
+D\left[\{s^z_1(\omega s^+_3+\omega^2 s_3^-)+(1\leftrightarrow 3)\}-\{s^z_2(\omega s^+_4+\omega^2 s^-_4)+(2\leftrightarrow 4)\}\right],
\label{eq_ptrans2}
\end{align}
\begin{align}
{P}_z^{(T)}=&L\left[\frac{1}{2}(s^x_1+s^x_2-s^x_3-s^x_4)+\frac{\sqrt{3}}{2}(s_1^y+s_2^y-s_3^y-s_4^y)\right]
+B\left[(s^+_1s^-_2+\text{h.c.})-(s^+_3s^-_4+\text{h.c.})\right]\nonumber\\
&+C\left[(s^+_1s^+_2+\text{h.c.})-(s^+_3s^+_4+\text{h.c.})\right]
+D\left[\{s^z_1(s^+_2+ s_2^-)+(1\leftrightarrow 2)\}-\{s^z_3(s^+_4+ s^-_4)+(3\leftrightarrow 4)\}\right],
\label{eq_ptrans3}
\end{align}
\end{widetext}
where $\omega=e^{i2\pi/3}$ and $B, C, D$, and $L$ are coupling constants which, just like $A$, depend on  microscopic details.

Time-reversal symmetry places additional constraints on the allowed terms in the polarization operator. For Kramers spin doublets (such as in Dy and Yb pyrochlores), $ \{ s^x, s^y, s^z \} \rightarrow - \{ s^x, s^y, s^z \}$ under time reversal, which enforces $L = 0$ by symmetry. However, for non-Kramers spin doublets (such as in Pr and Tb pyrochlores), $s^z\rightarrow -s^z$ and $\{s^x,s^y\}\rightarrow \{s^x,s^y\}$ under time reversal, because the transverse components are proportional to the magnetic quadrupole moment, whereas the longitudinal component is proportional to the magnetic dipole moment.\cite{lee2012generic} This enforces $D=0$ in the above expressions for the polarization operator.

Although these quantum terms have zero projection in the ice manifold as they generate spin flips, they can mediate virtual processes connecting different ground states and thereby generate non-trivial matrix elements within the classically degenerate ice manifold. This leads to non-trivial magnetoelectric effects much below the magnetic monopole excitation gap, as we show below.

The polarization operator for a down tetrahedra, in the presence of inversion symmetry, is given by replacing the above coupling constants as follows:
\begin{align}
\{A,B, C, D, L\}\rightarrow\{-A,-B,-C, -D, -L\}.
\end{align}
 The polarization operator thus has the right symmetry to couple to an external electric field. 
 
 Importantly, our derivation assumes that the pseudospin operators transform as pseudovectors under space group operations. These transformations are however expected to be more complicated in the case of `dipolar-octupolar' doublets, where one of the pseudospin components transforms as part of a magnetic octupolar tensor \cite{PhysRevLett.112.167203, PhysRevB.95.041106}. This physics is expected to be relevant in Nd and Ce-based pyrochlores, which also include promising QSL candidates \cite{PhysRevLett.115.097202, Petit2016}.

\subsection{Coupling to a uniform external electric field}
The coupling to a uniform external electric field ${\bf E}$ is given, to leading order, by
\begin{align}
H_{\bf E}=\alpha\sum_{\boxtimes}{\bf P}_\boxtimes\cdot{\bf E},
\label{eqn_polop}
\end{align}
where $\alpha$ is a coupling constant which allows to keep track of the perturbation theory (see below) and $\boxtimes$ denotes a sum over all tetrahedra. Hence the full spin Hamiltonian, in the presence of an electric field, is given by
\begin{align}
\mathcal{H}=H+H_{\bf E},
\label{eq_haminfield}
\end{align}
where $H$ is given by Eq. (\ref{eq_quantum spin iceham}). To simplify calculations illustrating the novel effects of the above magnetoelectric coupling, we concentrate on the case  $L=C=D=0$ in Eq. (\ref{eq_polop}). We shall discuss the effects of some of these couplings in due course.


\subsection{Effective low-energy Hamiltonian in the presence of an external electric field}

We now consider the regime where $J_{zz}\gg J_\pm, |{\bf E}|$. In this regime, we can perform perturbation theory to get the effective Hamiltonian in the presence of the electric field, to the leading non-zero order in $A$ and $B$, by extending the calculations of Hermele {\it et al.}\cite{PhysRevB.69.064404}. This is done in Appendix \ref{appen_pert1}. To the leading non-trivial (3rd) order, the effective Hamiltonian is given by (up to constants):
\begin{align}
\mathcal{H}_{\text{eff}}=& -g \sum_{{\hexagon}} \left(\mathcal{O}_{\hexagon}+\text{h.c.}\right)\nonumber\\
& - g \chi \sum_{m=1}^4  \left[3( \hat{\bf E}\cdot\hat{\bf t}_m)^2-1\right] \sum_{{\hexagon}\perp \hat{\bf t}_m}\left(\mathcal{O}_{\hexagon}+\text{h.c.}\right),
\label{eq_eff_ham1}
\end{align}
where $\chi \equiv \alpha^2B^2 \mathbf{E}^2/2 J_{\pm}^2$ is a dimensionless parameter, $\hat{\mathbf{E}}$ is the unit vector pointing in the direction of the external electric field, and $\mathcal{O}_{\hexagon}=s_1^+s_2^-s_3^+s_4^-s_5^+s_6^-$ ($1,\dots 6\in {\hexagon}$), $g = 12 J_\pm^3/J_{zz}^2$ as before. This expression reduces to Eq. (\ref{eq_eff_ham0}) when ${\bf E}=0$. As a side remark, we note that the above polarization also affects the dielectric properties of the system. As discussed in Appendix \ref{appen_pert1}, the coupling in Eq. (\ref{eqn_polop}) leads to additional contributions to the dielectric constant which can be of experimental interest, similar to that of classical spin ice.\cite{PhysRevB.72.144422}

Following the mapping described in Section \ref{sec_quantum spin iceham}, the  effective Hamiltonian [Eq. (\ref{eq_eff_ham1})] can be rewritten as
\begin{align}
\mathcal{H}_{\text{eff}}&=\frac{U}{2} \sum_{\left< \bf rr'\right>} { b}_{\bf rr'}^2- 2g \sum_{\left<\bf ss'\right>} \cos[{ e}_{\bf ss'}]\nonumber\\
&  -2g \chi \sum_{m=1}^4 \left[3 (\hat{{\bf E}} \cdot\hat{\bf t}_m)^2 - 1 \right]\sum_{\left< {\bf ss'}\right>\parallel \hat{\bf t}_m}\cos[{e}_{\bf ss'}],
\label{eq_lattice_ham}
\end{align}
This should be contrasted with Eq. (\ref{eq_lattice_ham0}). It is now clear that even in the presence of an external electric field, we have a pure compact $U(1)$ lattice gauge theory at low energies, albeit with anisotropic couplings. 

We now introduce a more convenient labeling convention.  Taking $\mathbf{r}$ and $\mathbf{s}$ to represent only up-tetrahedra, we denote the bonds on the direct and dual diamond lattices, respectively, by $({\bf r},n) \equiv (\mathbf{r}, \mathbf{r} + \mathbf{t}_n)$ and $({\bf s},m) \equiv (\mathbf{s}, \mathbf{s} + \mathbf{t}_m$), where $\mathbf{t}_n = \frac{a_0 \sqrt{3}}{4} \hat{\mathbf{t}}_n$ (similarly for $ \mathbf{t}_m$) and $a_0$ is the dimension of the cubic unit cell (see Appendix \ref{appen_pyrochlorelattice}). \footnote{Similarly, we label the bonds directed from down-tetrahedra to up-tetrahedra as $ (\mathbf{r},-n) \equiv (\mathbf{r} + \mathbf{t}_n, \mathbf{r})$ and $ (\mathbf{s},-m) \equiv (\mathbf{s} + \mathbf{t}_m, \mathbf{s})$. However, to avoid double-counting, these bonds do not appear in the Hamiltonian [Eq. (\ref{eq_lattice_ham_matrixform})]}. We thus rewrite Eq. (\ref{eq_lattice_ham}) as 
\begin{align}
\mathcal{H}_{\text{eff}}=&\frac{U}{2}\sum_{ {\bf r}, n} b_{ \mathbf{r}, n}^2 - \sum_{\mathbf{s},m} \mathcal{M}_m \cos[e_{\mathbf{s}, m}],
\label{eq_lattice_ham_matrixform}
\end{align}
where the coefficients $\mathcal{M}_m$ are given by
\begin{align}
\mathcal{M}_m &= 2g \left(1 - \chi + 3 \chi (\hat{\mathbf{E}} \cdot \hat{\mathbf{t}}_m)^2 \right).
\label{eq_Mmm}
\end{align}

Having derived the low-energy Hamiltonian in the presence of an external electric field, we now explain the implication of this coupling for the low-energy physics of quantum spin ice.  To this end, we note that while $J_{zz}\gg J_\pm, |{\bf E}|$ throughout this work, the relative strength of $J_\pm$ and $|{\bf E}|$, characterized by the dimensionless ratio $\chi=\alpha^2B^2{\bf E}^2/2J_\pm^2$, leads to two possible regimes: small electric fields (when $\chi \lesssim 1$) and large electric fields (when $\chi\gtrsim 1$). We discuss the physics of both these regimes in turn, starting with the small field limit. 

\section{Small field limit: Emergent birefringence}
\label{sec_biref}

 As discussed earlier, the ground state of the minimal quantum spin ice model, Eq. (\ref{eq_eff_ham0}), is a $U(1)$ QSL with a low-energy spectrum dominated by the gapless, linearly-dispersing emergent photons. The QSL is a deconfined phase of the gauge theory. One can neglect the gapped electric and magnetic charges at low energies, and hence neglect the compactness of the gauge group. Thus the cosine terms in the Hamiltonian in Eq. (\ref{eq_lattice_ham_matrixform}) can be expanded to give
\begin{align}
\mathcal{H}_{\text{eff}}=&\frac{U}{2}\sum_{ {\bf r}, n} b_{ \mathbf{r}, n}^2 + \frac{1}{2} \sum_{\mathbf{s},m} \mathcal{M}_m e_{\mathbf{s}, m}^2,
\label{eq_lattice_ham_matrixform_expanded}
\end{align}

We show next that, when subjected to a small, uniform external electric field ($\chi \lesssim 1$), the emergent photons acquire a birefringent behavior.
\subsection{Continuum theory}

The Hamiltonian in Eq. (\ref{eq_lattice_ham_matrixform_expanded}) is quadratic in the fields and can be diagonalized to find the dispersion of the excitations -- the emergent photons. However, before obtaining the full solutions on a lattice (see Sec. \ref{sec_latticetheory}), it is insightful to obtain the continuum theory which is valid for long wavelengths. Extending the prescription outlined in the previous section to the Hamiltonian in Eq. (\ref{eq_lattice_ham_matrixform_expanded}), we get
\begin{align}
\mathcal{H}_{\text{eff}}^{\rm continuum}=\frac{1}{2}\int d^3{\bf r}\left[\mathcal{U}~{\bf b}^2+e^\alpha\mathcal{K}^{\alpha\beta}e^\beta\right],
\label{eq_cont_ham}
\end{align}
where $\mathcal{U} = 4U/3l$, and the matrix $\mathcal{K}$ is given by
\begin{align}
\mathcal{K} = \mathcal{K}_0 \left[\begin{array}{ccc}
1 & 2\chi \hat{E}_x \hat{E}_y & 2\chi \hat{E}_x \hat{E}_z\\
2\chi \hat{E}_x \hat{E}_y & 1 & 2\chi \hat{E}_y \hat{E}_z\\
2\chi \hat{E}_x \hat{E}_z & 2\chi \hat{E}_y \hat{E}_z & 1 \\
\end{array}\right],
\label{eq_K_matrixform}
\end{align}
where $ \mathcal{K}_0 = 8g/3l$ as before, $\hat{E}_x \equiv \hat{\mathbf{E}} \cdot \hat{\mathbf{x}}$, and similarly for the $y$ and $z$ directions. [Einstein summation convention is assumed on all repeated Greek indices.] Comparing Eq. (\ref{eq_cont_ham}) with Eq. (\ref{eq_cont_ham0}), we note that the effect of the external electric field is to make the coupling of the emergent electric field non-diagonal while keeping the other basic features of the Maxwell theory intact.

The photon dispersion relations can now be obtained by diagonalizing $\mathcal{K}$. The eigenvalues of this matrix are positive definite for small values of the electric field (we consider the case of large electric fields in the next section) and hence using the Cholesky decomposition\cite{golub2012matrix} of symmetric positive-definite matrices, we can write
\begin{align}
\mathcal{K}=\Xi\cdot\Xi^T,
\end{align}
where $\Xi$ is a lower triangular matrix and $\Xi^T$ is its transpose. We introduce the vector (gauge) potential ${\bf A}$, which is related to the emergent fields using
\begin{align}
{\bf b}=\nabla\times {\bf A}, \quad e^\alpha= - \left[\mathcal{K}^{-1}\right]^{\alpha\beta}\partial_t A^\beta,
\label{eq_continuum_eandb}
\end{align}
and we quantize $\mathbf{A}$ as 
\begin{align}
A^\alpha (\mathbf{r})= \sum_{\lambda}\int \frac{d^3{\bf k}}{(2\pi)^3}  \frac{\Xi^{\alpha\beta}}{\sqrt{2\omega_\lambda(\mathbf{k})}} \Big[ e^{-i\mathbf{k} \cdot \mathbf{r}} {\mathbf{\epsilon}}^\beta_\lambda (\mathbf{k}) a_{\lambda \mathbf{k}} \nonumber \\
+ e^{i\mathbf{k} \cdot \mathbf{r}} \left\{{\mathbf{\epsilon}}^\beta_\lambda(\mathbf{k})\right\}^* a_{\lambda \mathbf{k}}^\dagger \Big],
\label{eq_continuum_A}
\end{align}
where $\omega_\lambda(\mathbf{k})$ and $\vect{\epsilon}_\lambda(\mathbf{k})$ are respectively the frequency and the unit polarization vector associated with polarization mode $\lambda$. From Eq. (\ref{eq_continuum_eandb}), the electric field is obtained:
\begin{align}
{e}^\alpha (\mathbf{r}) 
=i \sum_{\lambda} \int \frac{d^3{\bf k}}{(2\pi)^3} \sqrt{\frac{\omega_\lambda(\mathbf{k})}{2}}\left[(\Xi^T)^{-1}\right]^{\alpha\beta} \nonumber \\
\times \Big[ e^{-i\mathbf{k} \cdot \mathbf{r}} {\mathbf{\epsilon}}^\beta_\lambda (\mathbf{k}) a_{\lambda \mathbf{k}} - e^{i\mathbf{k} \cdot \mathbf{r}} \left\{{\mathbf{\epsilon}}^\beta_\lambda(\mathbf{k})\right\}^* a_{\lambda \mathbf{k}}^\dagger \Big],
\label{eq_continuum_e}
\end{align}	
where we used the Cholesky decomposition of $\mathcal{K}^{-1}$. Using Eq. (\ref{eq_continuum_eandb}), the magnetic field is given by
\begin{align}
b^\alpha({\bf r})&=\epsilon^{\alpha\beta\gamma}\partial_\beta A^\gamma({\bf r}) \nonumber \\
&=-i\epsilon^{\alpha\beta\gamma} \sum_{\lambda}\int \frac{d^3{\bf k}}{(2\pi)^3}  \frac{\Xi^{\gamma\rho}  k^\beta}{\sqrt{2\omega_\lambda(\mathbf{k})}} \Big[ e^{-i\mathbf{k} \cdot r}{\mathbf{\epsilon}}^\rho_\lambda (\mathbf{k}) a_{\lambda \mathbf{k}} \nonumber \\ &\hspace{3cm} - e^{i\mathbf{k} \cdot \mathbf{r}} \left\{{\mathbf{\epsilon}}^\rho_\lambda(\mathbf{k})\right\}^* a_{\lambda \mathbf{k}}^\dagger \Big].
\label{eq_continuum_b}
\end{align}
 We now impose that the bosonic operators satisfy the commutation relations $[a_{\lambda{\bf k}},a^\dagger_{\lambda'{\bf k'}}]= \left( 2 \pi \right)^3 \delta_{\lambda\lambda'}\delta_{\bf kk'}$. This gives rise to the appropriate commutation relations for the gauge potential and the electric field, {\it i.e.},
\begin{align}
\left[e^\alpha({\bf r}),A^\beta({\bf r'})\right]=i\delta_{\alpha\beta}\delta_{\bf rr'} .
\end{align}
Inserting the above expressions for the emergent electric and magnetic fields (Eqs. (\ref{eq_continuum_e}) and (\ref{eq_continuum_b}) respectively) into the continuum Hamiltonian in Eq. (\ref{eq_cont_ham}), we obtain the dispersion relation for the photons (see Appendix \ref{appen_continuum_photons}):
\begin{align}
\omega_\lambda(\mathbf{k}) = |\xi_\lambda(\hat{\mathbf{k}})| |\mathbf{k}|,
\label{eq_dispersion_continuum}
\end{align}
where $\xi_\lambda^2(\hat{\mathbf{k}})$ denote the eigenvalues of the Hermitian, positive-definite matrix
\begin{align}
Q(\hat{\mathbf{k}}) = \mathcal{U}~ \Xi^T \begin{pmatrix}
1 - \hat{k}_x^2     & -\hat{k}_x\hat{k}_y & -\hat{k}_x\hat{k}_z \\
-\hat{k}_x\hat{k}_y & 1 - \hat{k}_y^2     & -\hat{k}_y\hat{k}_z \\
-\hat{k}_x\hat{k}_z & -\hat{k}_y\hat{k}_z & 1 - \hat{k}_z^2
\end{pmatrix} \Xi
\end{align}
which only depends on the direction of $\mathbf{k}$ (not on its magnitude). Therefore, the photons are indeed gapless and linearly-dispersing, as expected from gauge invariance.

However, the speed of emergent light depends on both the direction of propagation -- encoded in $Q(\hat{ \mathbf{k} })$ -- and the polarization mode $\lambda$, as shown in Fig. \ref{fig_dispersion} for electric fields in the $[011]$ and $[111]$ directions. This is very similar to the physics of birefringent materials; however, we stress that the photons discussed here are emergent and do not exist outside the material. As expected from $U(1)$ gauge invariance, one of the three polarization modes obtained from diagonalizing $Q(\hat{\mathbf{k}})$, that always has zero energy, is unphysical and does not couple to any observable. This leaves us with only the two familiar propagating modes, that are in general not transverse to the propagation direction $\mathbf{k}$ of the photons -- again in analogy with birefringent materials. Somewhat similar birefringent behavior of emergent photons was recently suggested in the context of hexagonal water ice.\cite{PhysRevB.93.125143} However, in contrast to the present case, there the birefringence stems from the inherent layered structure of the system which gives rise to inequivalent directions.

We note that electric fields applied along the crystallographic cubic axes ($[100]$, $[010]$ or $[001]$) lead to a diagonal matrix $\mathcal{K}$ (see Eq. (\ref{eq_K_matrixform})), and thus do not produce any birefringence.  For all other directions, the external electric field leads to birefringence for the emergent photons.
\subsection{Lattice theory}
\label{sec_latticetheory}

\begin{figure*}
	\centering
	\includegraphics[width=0.49\textwidth]{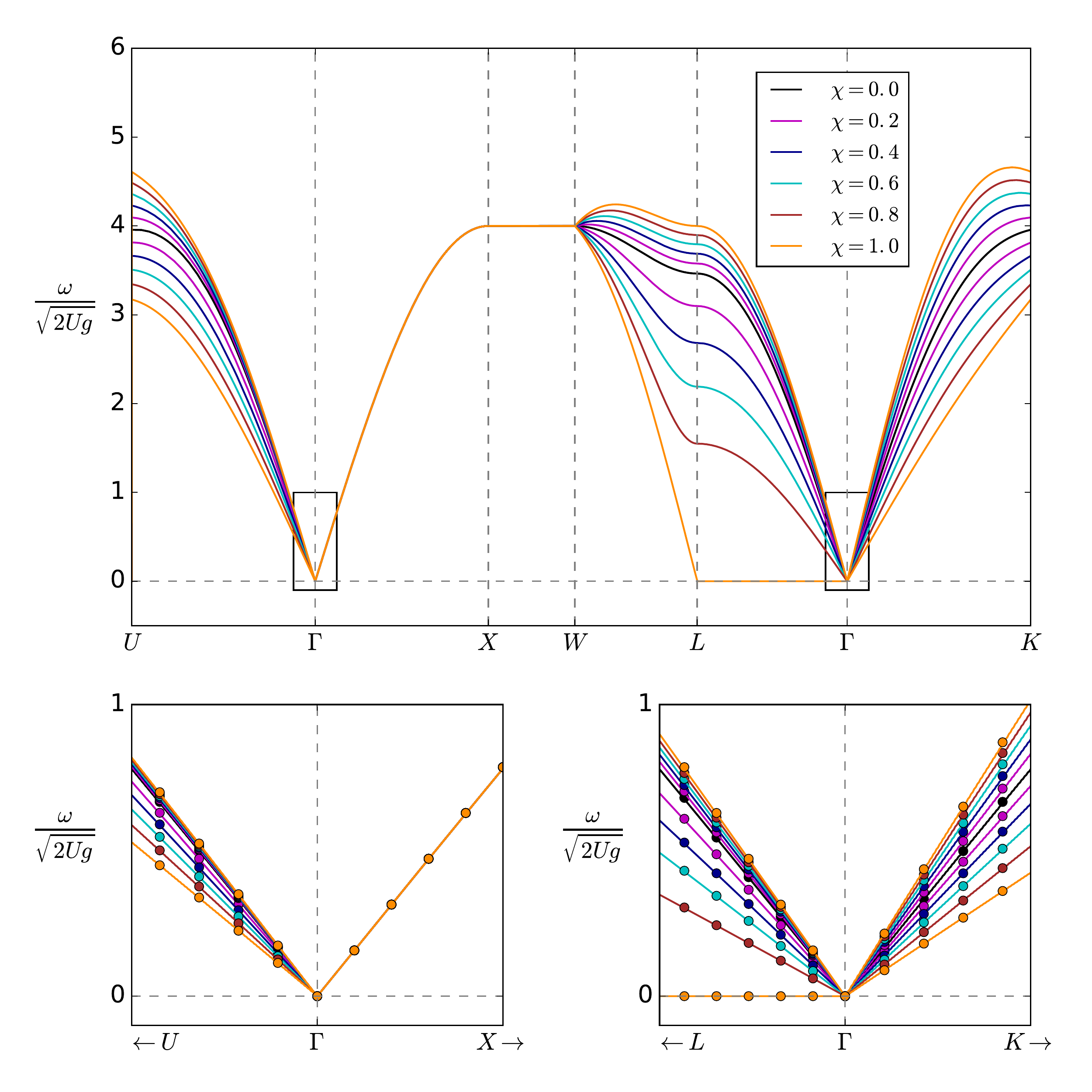}
	\includegraphics[width=0.49\textwidth]{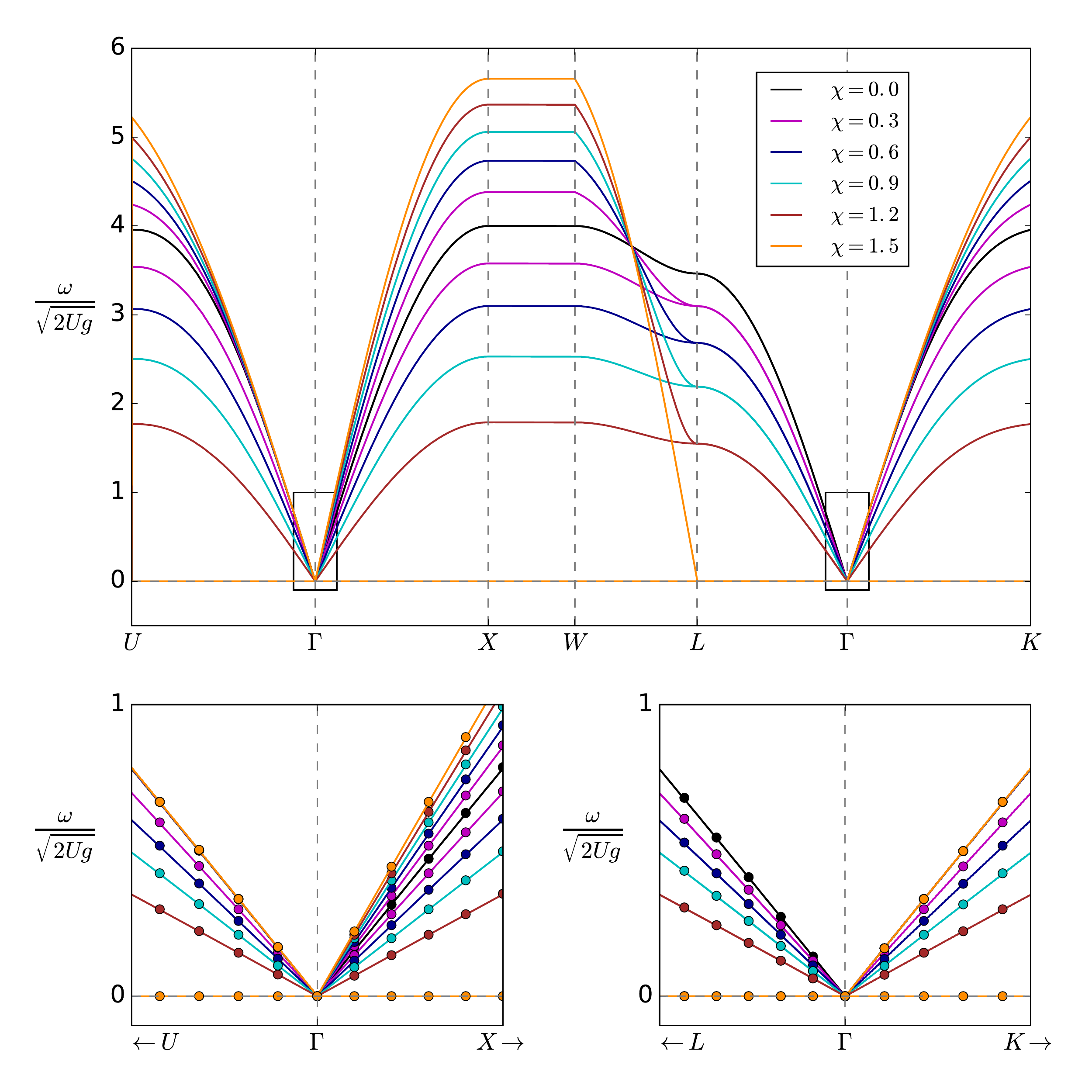}
	\caption{Frequency of the emergent photons under a uniform electric field along the $[011]$ direction (left) and the $[111]$ direction (right). The two physical polarization modes are shown, for applied electric field strengths $\chi$ in the low field regime. Upper panels: frequencies in the lattice theory, as obtained from Eq. (\ref{eq_lattice_dispersion}). Lower panels: the solid lines represent the lattice theory, and the circles come from the continuum theory [Eq. (\ref{eq_dispersion_continuum})], in units of  $\sqrt{\mathcal{U} \mathcal{K}_0}$), the latter being valid only near the center of the Brillouin zone. The labels on the horizontal axis correspond to the following high-symmetry points of the FCC lattice: $U = [\pi/2,2 \pi, \pi/2]$, $X = [0,2\pi,0]$,  $W = [\pi,2\pi,0]$, $L = [\pi,\pi,\pi]$ and $K = [3\pi/2,3\pi/2,0]$, with the cubic lattice constant set to $a_0=1$.}
	\label{fig_dispersion}
\end{figure*}
The continuum theory described above is valid only for long wavelengths, {\it i.e.}, near the Brillouin zone center. We now turn to the full lattice theory of emergent electromagnetism in quantum spin ice under a small, uniform external electric field.  To this end, we generalize the approach taken by Benton \textit{et al.} \cite{PhysRevB.86.075154} but starting with the lattice Hamiltonian in Eq. (\ref{eq_lattice_ham_matrixform_expanded}). 
As in the continuum theory, we introduce a vector (gauge) potential $A$, defined on the bonds of the dual diamond lattice as
	\begin{align}
	A_{(\mathbf{s},m)} =& \sqrt{\frac{2}{N}} \sum_{\mathbf{k}, \lambda} \sqrt{\frac{\mathcal{M}_m}{\omega_\lambda(\mathbf{k})}} \Big[ e^{-i\mathbf{k} \cdot (\mathbf{s} + \mathbf{t}_m/2)} \eta_{m \lambda} (\mathbf{k}) a_{\lambda \mathbf{k}} \nonumber\\
	 &~~~~~~~~~~~~~~~~~~+ e^{i\mathbf{k} \cdot (\mathbf{s} + \mathbf{t}_m/2)} \eta^*_{m \lambda}(\mathbf{k}) a_{\lambda \mathbf{k}}^\dagger \Big],
	\label{eq_lattice_a}
	\end{align}
	where $m=1-4$, $N$ is the number of spins in the system, and the 4-component vectors $\vect{\eta}_\lambda$ are analogous to the polarization vectors $\vect{\epsilon}_\lambda$ of the continuum theory, but expressed in the sublattice ($\hat{\mathbf{t}}_m$) basis. The electric field is defined as
	\begin{align}
	e_{(\mathbf{s},m)} =& -\frac{1}{\mathcal{M}_m} \frac{\partial A_{(\mathbf{s},m)}}{\partial t} \nonumber \\
	=&i \sqrt{\frac{2}{N}} \sum_{\mathbf{k},\lambda} \sqrt{\frac{\omega_\lambda(\mathbf{k})}{\mathcal{M}_m}} \Big[ e^{-i\mathbf{k} \cdot (\mathbf{s} + \mathbf{t}_m/2)} \eta_{m \lambda} (\mathbf{k}) a_{\lambda \mathbf{k}} \nonumber \\
	&~~~~~~~~~~~~~~~~~~~- e^{i\mathbf{k} \cdot (\mathbf{s} + \mathbf{t}_m/2)} \eta^*_{m \lambda}(\mathbf{k}) a_{\lambda \mathbf{k}}^\dagger \Big].
	\label{eq_lattice_e}
	\end{align}
The emergent fields $A$ and $e$ are directed variables on the dual diamond lattice, i.e., $A_{(\mathbf{s},-m)} = -A_{(\mathbf{s},m)}$ and  $e_{(\mathbf{s},-m)} = -e_{(\mathbf{s},m)} $. Enforcing the bosonic commutation relations $[a_{\lambda{\bf k}},a^\dagger_{\lambda'{\bf k'}}]= \delta_{\lambda\lambda'}\delta_{\bf kk'} $ leads to the correct commutation relations from electromagnetism, which on the lattice are given by 
\begin{align}
[e_{(\mathbf{s},m)},A_{(\mathbf{s'},m')}] = i \delta_{\mathbf{s}\mathbf{s'}} ( \delta_{mm'} - \delta_{m,-m'}).
\label{eq_lattice_commutationrelations}
\end{align}

The magnetic field $b$ is obtained by taking the lattice curl of the gauge field $A_{(\mathbf{s},m)}$ on the dual diamond lattice,
\begin{align}
b_{(\mathbf{r},n)} =(\nabla_{\hexagon} \times A_{(\mathbf{s},m)})_{(\mathbf{r},n)}.
\label{eq_lattice_ba}
\end{align}
The details of the computation of the lattice curl are given in Appendix \ref{appen_lattice_photons}. We get:
\begin{align}
b_{(\mathbf{r},n)} =& \sqrt{\frac{2}{N}} \sum_{\mathbf{k}, \lambda, m} \sqrt{\frac{\mathcal{M}_m}{\omega_\lambda(\mathbf{k})}} \Big[ e^{-i\mathbf{k} \cdot (\mathbf{r} - \mathbf{t}_n/2)}  Z_{nm}(\mathbf{k}) \eta_{m \lambda}(\mathbf{k}) a_{\lambda \mathbf{k}}  \nonumber \\
&+ e^{i\mathbf{k} \cdot (\mathbf{r} - \mathbf{t}_n/2)} Z_{mn}(\mathbf{k}) \eta_{m \lambda}^*(\mathbf{k})   a_{\lambda \mathbf{k}}^\dagger \Big],
\label{eq_lattice_b}
\end{align}
where, following Ref. \onlinecite{PhysRevB.86.075154}, we defined the Hermitian matrix
\begin{align}
Z(\mathbf{k}) = -2i \begin{pmatrix} 0 & s_{01}(\mathbf{k}) & s_{02}(\mathbf{k}) & s_{03}(\mathbf{k}) \\
-s_{01}(\mathbf{k}) & 0 & s_{12}(\mathbf{k}) & s_{13}(\mathbf{k}) \\
-s_{02}(\mathbf{k}) & -s_{12}(\mathbf{k}) & 0 & s_{23}(\mathbf{k}) \\
-s_{03}(\mathbf{k}) & -s_{13}(\mathbf{k}) & -s_{23}(\mathbf{k}) & 0 \end{pmatrix},
\label{eq_Zmatrix_definition_maintext}
\end{align}
with $$s_{nm}(\mathbf{k}) \equiv \sin \left( \mathbf{k} \cdot \frac{a_0}{\sqrt{8}} \frac{\hat{\mathbf{t}}_n \times \hat{\mathbf{t}}_m }{ |\hat{\mathbf{t}}_n \times \hat{\mathbf{t}}_m |} \right),$$
Inserting Eqs. (\ref{eq_lattice_e}) and (\ref{eq_lattice_b}) for the emergent fields in the lattice Hamiltonian [Eq. (\ref{eq_lattice_ham_matrixform_expanded})], we arrive at the following dispersion relation for the photons (see Appendix \ref{appen_lattice_photons}):
\begin{align}
\omega_\lambda (\mathbf{k}) =  |\zeta_\lambda(\mathbf{k})|,
\label{eq_lattice_dispersion}
\end{align}
where $\zeta^2_\lambda(\mathbf{k})$ are the eigenvalues of the Hermitian, positive-definite matrix $T(\mathbf{k})$ with elements given by
\begin{align}
T_{m'm}(\mathbf{k}) =  U \sqrt{\mathcal{M}_{m'} \mathcal{M}_m} [Z^2(\mathbf{k})]_{m'm}  .
\label{eq_Tmatrix}
\end{align}

We present in Fig. \ref{fig_dispersion} the photon dispersion relations obtained for electric fields of various strengths in the $[011]$ and $[111]$ directions, using Eq. (\ref{eq_lattice_dispersion}). 

As expected from $U(1)$ gauge invariance, we always obtain only two physical (non-zero energy) polarization modes. The speed of emergent light becomes direction- and polarization-dependent (for $\chi \neq 0$), and the low-energy limit of the lattice theory [Eq. (\ref{eq_lattice_dispersion})] reproduces the continuum calculations [Eq. (\ref{eq_dispersion_continuum})] near the Brillouin zone center.  We comment here on two striking features. Firstly, some propagation directions are unaffected by the electric field -- \textit{e.g.}, the $\Gamma - X$ and $X - W$ bands for $\mathbf{E} \propto [011]$ -- because of the symmetries of the pyrochlore lattice under the applied field. Secondly, for a critical value of the field ($\chi = 1$ for the $[011]$ direction, and $\chi = 1.5$ for the $[111]$ direction), the speed of the emergent photons vanishes for some wavevectors and polarization modes. We thus expect instabilities to develop and the nature of the ground state to change, as explained in section \ref{sec_symmetrybreaking}.
\subsection{Spin structure factors}

Having obtained the photon dispersion relations, we now investigate the spin structure factors, which are measured in neutron scattering experiments, in the presence of an external electric field. Such measurements are already quite common in the context of multiferroics.

The \textit{energy-integrated} (or \textit{equal-time}) spin structure factor is given by
\begin{align}
I^{\alpha \beta} (\mathbf{k}, t=0) =  \int d\omega~ I^{\alpha \beta} (\mathbf{k}, \omega), 
\label{eq_et_sf}
\end{align}
where $I^{\alpha \beta} (\mathbf{k}, \omega)$ is the dynamic spin structure factor:
\begin{align}
I^{\alpha \beta} (\mathbf{k}, \omega) =  \int dt~e^{-i\omega t} \left< s^\alpha(-\mathbf{k},t) s^\beta(\mathbf{k},0)\right>.	
\label{eq_dy_sf}
\end{align}
We use the mapping of spins $\mathbf{s}$ onto the emergent lattice magnetic field $b$ (given by Eq. \ref{eq_lattice_b}), projecting the corresponding spin quantization axis $\hat{\mathbf{t}}_m$ ($\hat{\mathbf{t}}_n$) onto the axis of interest represented by the unit vector $\hat{\mathbf{\alpha}}$ ($\hat{\mathbf{\beta}}$) \cite{PhysRevB.86.075154}}:

	\begin{align}
	\left< s^\alpha(-\mathbf{k},t) s^\beta(\mathbf{k},0) \right> =& \sum_{m,n} (\hat{\mathbf{t}}_m\cdot\hat{\alpha})(\hat{\mathbf{t}}_n\cdot\hat{\beta})\nonumber\\
	& \times \left< b_m(-\mathbf{k},t) b_n(\mathbf{k},0)\right>,
	\end{align}
We find that the equal-time (zero-temperature) structure factor becomes, for $\chi \lesssim 1$ (see Appendix \ref{appen_structurefactors}):
\begin{align}
I&^{\alpha \beta} (\mathbf{k}, t=0) =  \frac{1}{8}\sum_{\lambda,m,n,l,l'} \frac{(\hat{\mathbf{t}}_m\cdot\hat{\alpha})(\hat{\mathbf{t}}_n\cdot\hat{\beta})}{ \omega_\lambda(\mathbf{k})}  \nonumber \\
&\times\eta^*_{l' \lambda}(\mathbf{k}) \sqrt{\mathcal{M}_{l'}} Z_{l'n}(\mathbf{k})  Z_{ml}(\mathbf{k}) \sqrt{\mathcal{M}_l} \eta_{l \lambda}(\mathbf{k}).
\label{eq_structurefactors_0}
\end{align} 

Following the convention used in Ref. \onlinecite{Fennell16102009} for a polarized neutron scattering experiment, we define the following coordinate system:
$$ \mathbf{x} \parallel \mathbf{k} , \quad \mathbf{y} \parallel \vect{\eta}_\nu \times \mathbf{k} , \quad \mathbf{z} \parallel \vect{\eta}_\nu ,$$
where $\vect{\eta}_\nu \perp \mathbf{k}$ is the neutron polarization direction, and we specialize to the spin-flip channel, where $ \hat{\alpha} = \hat{\beta} = \hat{ \mathbf{y} }$. 
Thus, Eq. (\ref{eq_structurefactors_0}) is rewritten as
\begin{align}
I&^{yy} (\mathbf{k}, t=0) =  \frac{1}{8}\sum_{\lambda,m,n,l,l'} \frac{\left( \hat{\mathbf{t}}_m \cdot \hat{\mathbf{y}} \right) \left( \hat{\mathbf{t}}_n \cdot \hat{\mathbf{y}} \right) } { \omega_\lambda(\mathbf{k})}  \nonumber \\
&\times\eta^*_{l' \lambda}(\mathbf{k}) \sqrt{\mathcal{M}_{l'}} Z_{l'n}(\mathbf{k})  Z_{ml}(\mathbf{k}) \sqrt{\mathcal{M}_l} \eta_{l \lambda}(\mathbf{k}),
\label{eq_structurefactors}
\end{align}

We plot the equal-time structure factor in the spin-flip channel in Fig. \ref{fig_structurefactors} for two directions of the external electric field: $[011]$ and $[111]$. Clearly, as the external electric field is increased, the structure factors evolve differently depending on the direction of the former. Indeed, the changes are more prominent for the direction [111], compared to the direction [011]. This can possibly be detected in experiments (see Sec. \ref{sec_materials} for an estimate of the field strengths required). 

As mentioned in the caption of Fig. \ref{fig_structurefactors}, the intensity of each of the plots is chosen independently. This is because the above expression, Eq. (\ref{eq_structurefactors}), is correct up to an anisotropic  renormalization factor which arises upon integration of the high-energy modes. We shall discuss the issue of the anisotropic renormalization of the structure factor intensities in Sec. \ref{sec_transition}. However, here we note that such effects are not expected to change the qualitative nature of the results shown in Fig. \ref{fig_structurefactors}.

\section{Large field limit: quantum spin liquids with emergent electric $\pi$-fluxes}
\label{sec_symmetrybreaking}

In Section \ref{sec_biref}, we described the small electric field regime which leads to birefringent behavior for the emergent photons. However, another interesting effect occurs when the field-induced terms in Eq. (\ref{eq_lattice_ham_matrixform}) become comparable to the usual third-order perturbation term, {\it i.e.}, $\chi \gtrsim 1$. In this section, we show that instabilities then develop, signaled by the effective velocity of the photons vanishing in some directions. This happens when at least one coupling constant $\mathcal{M}_m$ of the emergent electric field (see Eq. (\ref{eq_lattice_ham_matrixform})) changes sign. From Eq. (\ref{eq_Mmm}), we see that this occurs when
\begin{align}
1 - 3(\hat{\mathbf{E}} \cdot \hat{\mathbf{t}}_m)^2 > \frac{1}{\chi}
\label{eq_condition_flux}
\end{align} 
for some $m$, {\it i.e.}, some orientation of the hexagons. Then, the coupling to the electric field term, $\cos[ e_{\mathbf{s}, m}]$, becomes \textit{positive} for all hexagons perpendicular to $\hat{\mathbf{t}}_m$. At the mean-field level, the energy is minimized for $e_{\mathbf{s},m} \sim \pi$, \textit{i.e.}, the corresponding hexagons now trap a background electric flux of $\pi$ around which all fluctuations occur. Such trapping of $\pi$-fluxes in quantum spin ice was first described in Ref. \onlinecite{lee2012generic} for a model with frustrated ($J_\pm < 0$) transverse exchange interactions. In contrast, here only a \textit{subset} of the hexagons trap a $\pi$-flux, in a way which depends sensitively on the applied electric field direction. 

 Exploiting the fact that $\cos[e_{\mathbf{s}, m} ] = - \cos[e_{\mathbf{s}, m} - \pi ]$,
 the low-energy expansion in this situation is obtained as follows. We rewrite Eq. (\ref{eq_lattice_ham_matrixform}) as 
\begin{align}
\mathcal{H}_{\text{eff}}=&\frac{U }{2}\sum_{ \mathbf{r}, n} b_{ \mathbf{r}, n}^2 - \sum_{\mathbf{s},m}  |\mathcal{M}_m| \cos[\tilde e_{\mathbf{s}, m}],
\label{eq_lattice_ham_compact}
\end{align}
where
\begin{align}
\tilde{e}_{\mathbf{s},m} = \begin{cases} &e_{\mathbf{s},m} - \pi \quad \: , \quad  \mathcal{M}_m < 0 \\
                         &e_{\mathbf{s},m} \qquad \quad , \quad \mathcal{M}_m > 0
                         \end{cases}
\end{align}
and, expanding the cosine term, we get
\begin{align}
\mathcal{H}_{\text{eff}}=&\frac{U}{2}\sum_{ \mathbf{r}, n}   b_{ \mathbf{r}, n}^2  + \frac{1}{2} \sum_{\mathbf{s},m} |\mathcal{M}_m| \tilde{e}_{\mathbf{s}, m}^2,
\label{eq_lattice_ham_reversedsign}
\end{align}
which is the same Hamiltonian as in Eq. (\ref{eq_lattice_ham_matrixform_expanded}), but with a background electric field of $\pi$ for a \textit{subset} of the hexagons.  

\begin{figure*}
	\subfigure[]{
		\includegraphics[width=0.58\columnwidth]{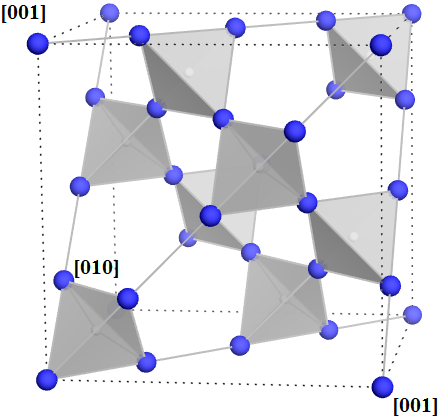}
		\label{fig_fluxpatterns_001}
	}
	\hspace{0.8cm}
	\subfigure[]{
		\includegraphics[width=0.58\columnwidth]{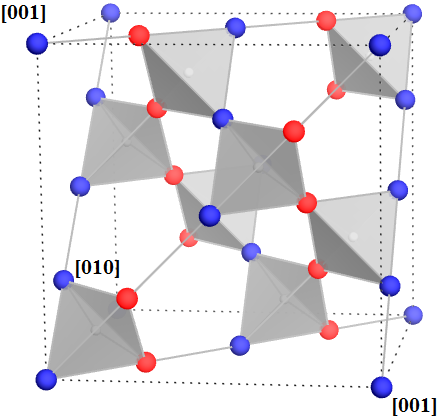}
		\label{fig_fluxpatterns_011}
	}
	\hspace{0.8cm}
	\subfigure[]{
		\includegraphics[width=0.58\columnwidth]{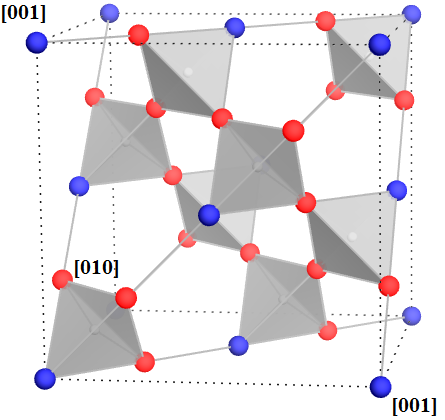}
		\label{fig_fluxpatterns_111}	
	}
	\caption{Schematic representation of the `flux patterns' on the dual pyrochlore lattice. Red sites are located at the center of flux-carrying hexagons on the direct pyrochlore lattice, while blue sites correspond to hexagons without a flux. (a): For $\mathbf{E} \propto [001]$, no fluxes are trapped. (b): For $\mathbf{E} \propto [011]$, flux-carrying sites form lines along $[0\bar{1}1]$  (perpendicular to the applied field). (b): For $\mathbf{E} \propto [111]$, flux-carrying sites form kagome planes perpendicular to the applied field.}
	\label{fig_fluxpatterns}
\end{figure*}

Below we describe three phases obtained for three different directions of the external electric field. Note that each of these phases has gapless emergent photons.  The gapped, but deconfined,   magnetic monopole excitations would see the electric fluxes as sources of Aharonov-Bohm phase and hence their band structure would change compared to the zero electric flux state. In other words, in the background of $\pi$-electric flux, the monopoles would transform under a different projective representation of the symmetry group of the system.  Also, the state does not break time-reversal symmetry as $\pi \rightarrow-\pi$ under time reversal. As a result, the total (emergent) electric charge in the $\pi$-flux states is zero. All these suggest that the present $U(1)$ QSLs are different from the zero-flux $U(1)$ QSL and are characterized by a different projective symmetry group. The comparative differences in the monopole band structure can have experimental consequences, {\it e.g.}, in the finite energy part of the spin structure factor or in possible quantum phase transitions brought about by the condensation of such monopoles.  
\subsection{${\bf E} \propto [001]$}
For an electric field along one of the crystallographic axes, $\mathcal{M}_m=1$ and the phase transition does not occur. This depicted in Fig. \ref{fig_fluxpatterns_001}.
\subsection{${\bf E} \propto [011]$} 

For an electric field in the [011] crystallographic direction, we have 
$\hat{\mathbf{E}} = 1/\sqrt{2} \{0,1,1\}$, and 
\begin{align*}
(\hat{\mathbf{E}} \cdot \hat{\mathbf{t}}_{1(4)})^2 =  2/3, \quad (\hat{\mathbf{E}} \cdot \hat{\mathbf{t}}_{2(3)})^2 = 0. 
\end{align*} 

Using Eq. (\ref{eq_condition_flux}), we see that the hexagons oriented along $\hat{\bf t}_1$ and $\hat {\bf t}_4$ never trap a flux, while those oriented along $\hat{\bf t}_2$ and $\hat {\bf t}_3$ do so  for $\chi > 1$. Since the hexagon centers on the direct pyrochlore lattice correspond to sites of the dual pyrochlore lattice, the flux lines are best viewed in the latter. In this dual lattice, they correspond to one-dimensional strings of electric flux running along $[0\bar{1}1]$ directions, as shown in Fig. \ref{fig_fluxpatterns_011}.
 
\subsection{${\bf E} \propto [111]$}
For an electric field in the [111] crystallographic direction, we have $\hat{\mathbf{E}} = 1/\sqrt{3} \left( 1,1,1 \right)$, and 
\begin{align*}
(\hat{\mathbf{E}} \cdot \hat{\mathbf{t}}_1)^2 = 1 \enspace, \quad (\hat{\mathbf{E}} \cdot \hat{\mathbf{t}}_{2(3,4)})^2  = \frac{1}{9}.
\end{align*}

Using Eq. (\ref{eq_condition_flux}), we see that the hexagons oriented along $\hat{\bf t}_1$ never trap a flux, but those oriented along  $\hat{\bf t}_2, \hat{\bf t}_3$ and $\hat{\bf t}_4$ trap fluxes for $\chi > 3/2$. As shown in Fig. \ref{fig_fluxpatterns_111}, in the dual pyrochlore lattice, this results in a `flux pattern' of parallel kagome planes of sites with fluxes, and triangular planes of sites without fluxes. 

\subsection{Other possibilities}

One can also obtain kagome planes of sites \textit{not} trapping a flux, and triangular planes of sites trapping a flux (see Fig. \ref{fig_fluxpatterns_111}, with blue and red colors inverted). This phase is achieved, \textit{e.g.}, by choosing an electric field in the $[\bar{2}11]$ direction and $1 < \chi < 3$.  We also note that, given Eq. (\ref{eq_condition_flux}), it is impossible to find a direction $\hat{\mathbf{E}}$ such that \textit{all} sublattices trap a flux.

\section{Renormalized spin structure factor and re-emergence of pinch points near the phase transition}
\label{sec_transition}

It is now interesting to ask about the nature of the phase transition between the $U(1)$ QSLs without (at low external electric field) and with (at high external electric field) emergent electric $\pi$-fluxes. It is clear from the discussion above that this transition is brought about by the condensation of emergent electric flux lines, the detailed pattern of which is dictated by the direction of the external field. The discontinuous jump of the trapped electric flux from $0$ to $\pi$ through some hexagons indicates that the transition is possibly first order.  Even then, we expect such a Landau-forbidden transition between two long-range entangled states of matter to be quite interesting\cite{PhysRevB.94.125112} -- however, we leave its investigation for a separate study.

In this section, we comment on the re-emergence of a set of {\it pinch points} in the spin structure factor, characteristic of classical spin ice,\cite{castelnovo2012spin, PhysRevB.23.232, PhysRevLett.93.167204, PhysRevB.71.014424} near the phase transition.  To understand this, we return to Eq. (\ref{eq_lattice_ham_matrixform_expanded}) and approach the transition from the low external electric field (zero-flux) side. As the photon velocity, given by Eq. (\ref{eq_lattice_dispersion}), vanishes at the transition (for certain directions), zero modes are generated and the theory appears unstable. To remedy this, we need to consider the next-to-leading order in perturbation theory that generates a dispersion for the emergent photons. Following Benton {\it et. al.},\cite{PhysRevB.86.075154} this is given by (also see Appendix \ref{appen_reg}): 
\begin{align}
\mathcal{H}_{\text{eff}}=&\frac{U}{2}\sum_{ {\bf r}, n} b_{ \mathbf{r}, n}^2 + \frac{1}{2} \sum_{\mathbf{s},m} \left[ \mathcal{M}_m e_{\mathbf{s}, m}^2 + \mathcal{W}_m (\nabla \times b)^2_{\mathbf{s},m} \right],
\label{eq_lattice_ham_matrixform_expanded_higherorder}
\end{align}
where the couplings $\mathcal{W}_m (>0)$ are direction-dependent, and can in principle be obtained from higher-order terms in perturbation theory, similarly to what is presented in detail in Appendix \ref{appen_pert1}.  This direction dependence is however not crucial for the regularization procedure, because the $\mathcal{W}_m$ coefficients do not vanish for the same electric field strength as the leading-order coefficients $\mathcal{M}_m$. Hence we take $\mathcal{W}_m=\mathcal{W}>0$ (say) for all directions. The regularized theory [Eq. (\ref{eq_lattice_ham_matrixform_expanded_higherorder})] then leads to a corrected dispersion relation for the photons given by (see Appendix \ref{appen_reg}):
\begin{align}
\omega_\lambda (\mathbf{k}) =  |\zeta_\lambda(\mathbf{k})|,
\label{eq_dispersion_reg}
\end{align}
where $\zeta^2_\lambda(\mathbf{k})$ are the eigenvalues of the Hermitian, positive-definite matrix 
\begin{align}
T(\mathbf{k}) + \mathcal{W} \left[Z(\mathbf{k})\right]^4
\label{eq_dispersion_reg_2} 
\end{align}
where, as before, $T({\bf k})$ and $Z({\bf k})$ are given by Eq. (\ref{eq_Tmatrix}) and (\ref{eq_Zmatrix_definition_maintext}), respectively.

\begin{figure}
	\centering
	\includegraphics[width=0.49\columnwidth]{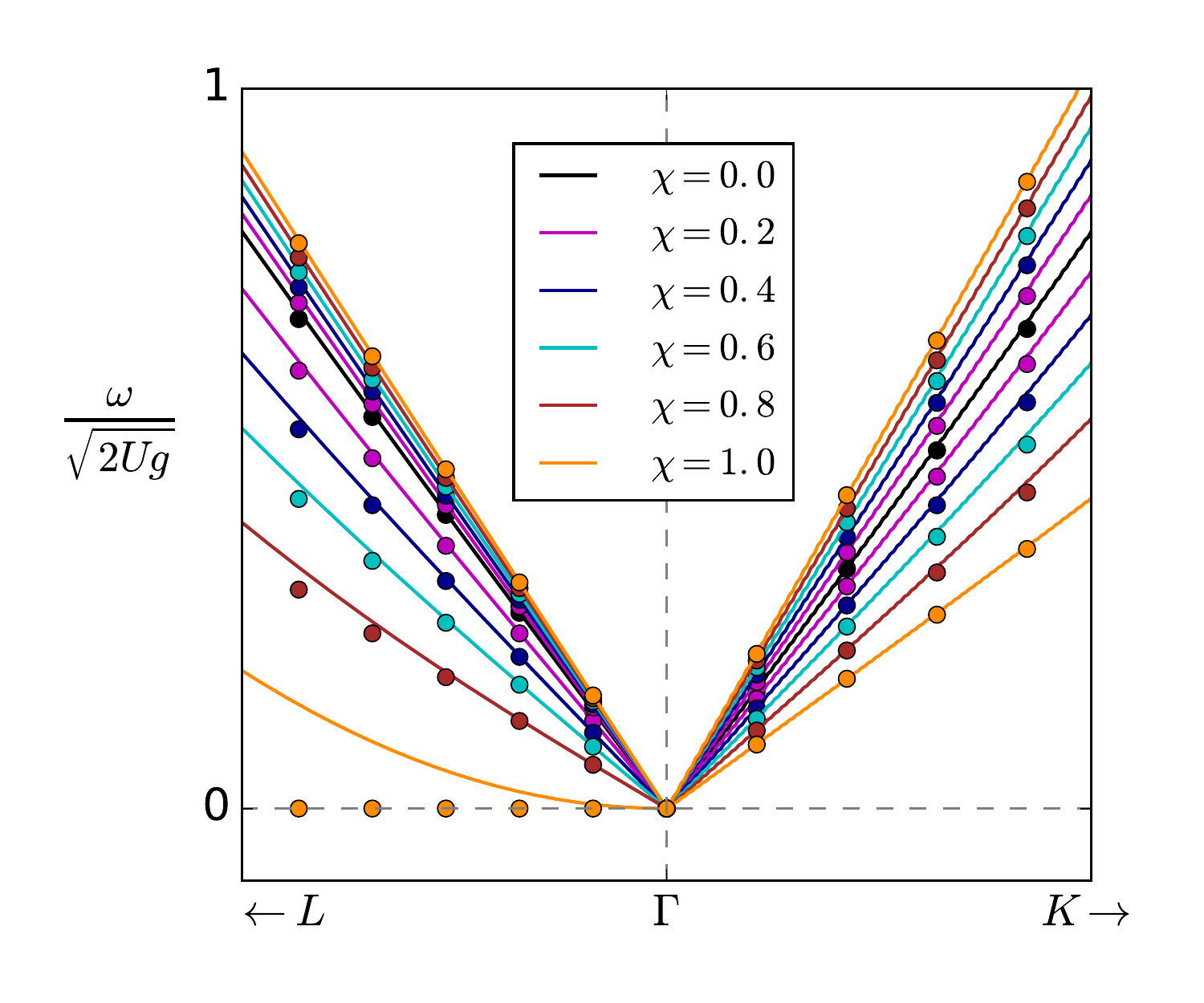}
	\includegraphics[width=0.49\columnwidth]{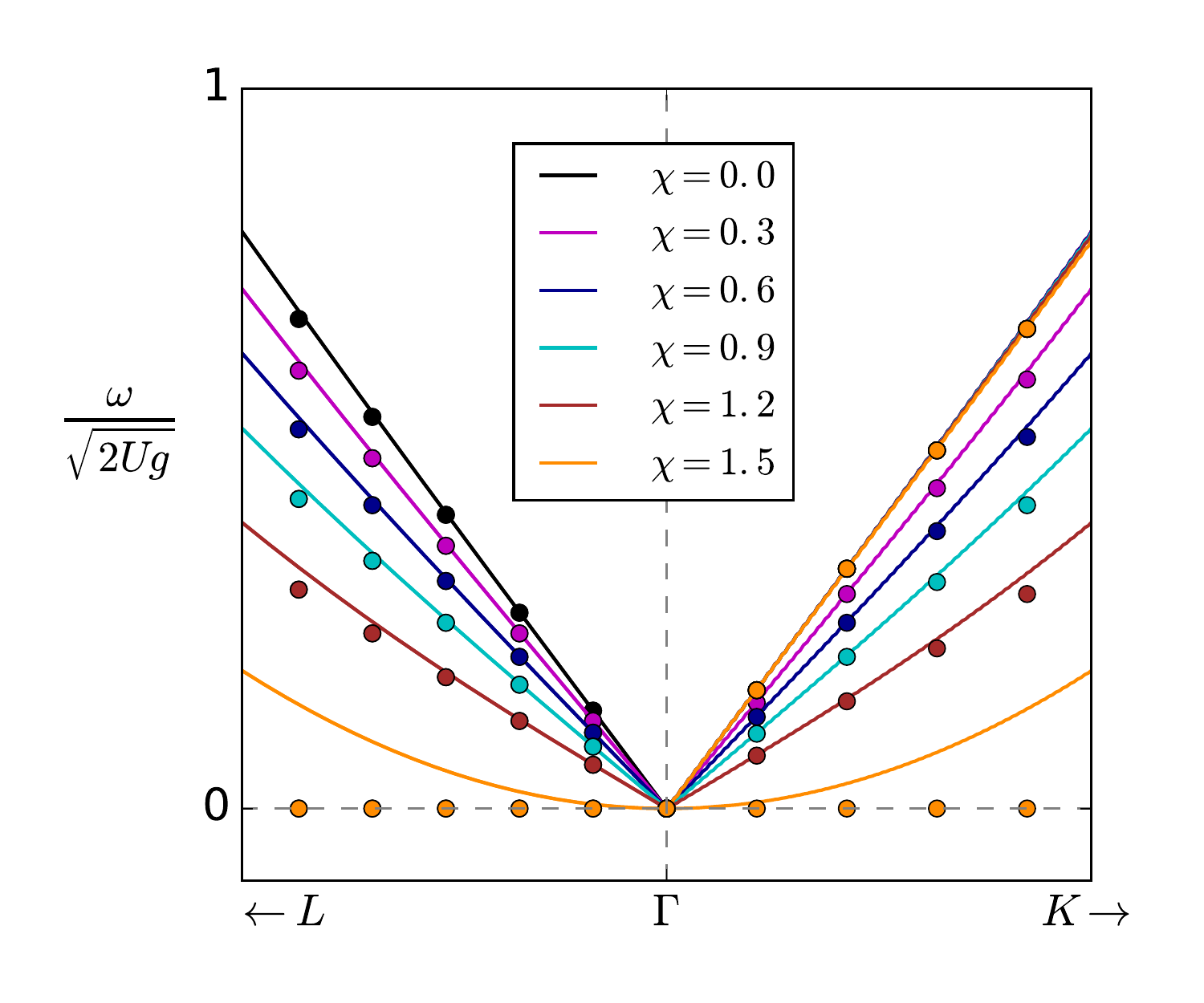}
	\caption{Photon spectrum near the $\Gamma$ point for an electric field along the $[011]$ direction (left) and the $[111]$ direction (right), regularized using $\mathcal{W}/2Ug = 0.1$. The solid lines represent the regularized lattice theory [Eq. (\ref{eq_dispersion_reg})], and the circles represent the unregularized lattice theory [Eq. (\ref{eq_lattice_dispersion})] for reference.}
	\label{fig_lattice_reg}
\end{figure}

As discussed in Ref. \onlinecite{PhysRevB.86.075154}, the effect of this new term is to endow the photons with a quadratic dispersion when the speed of light vanishes, as shown in Fig. \ref{fig_lattice_reg} for  external electric fields ranging from 0 up to the critical point.

\begin{figure}[h]
	\centering
	\includegraphics[width=0.9\columnwidth]{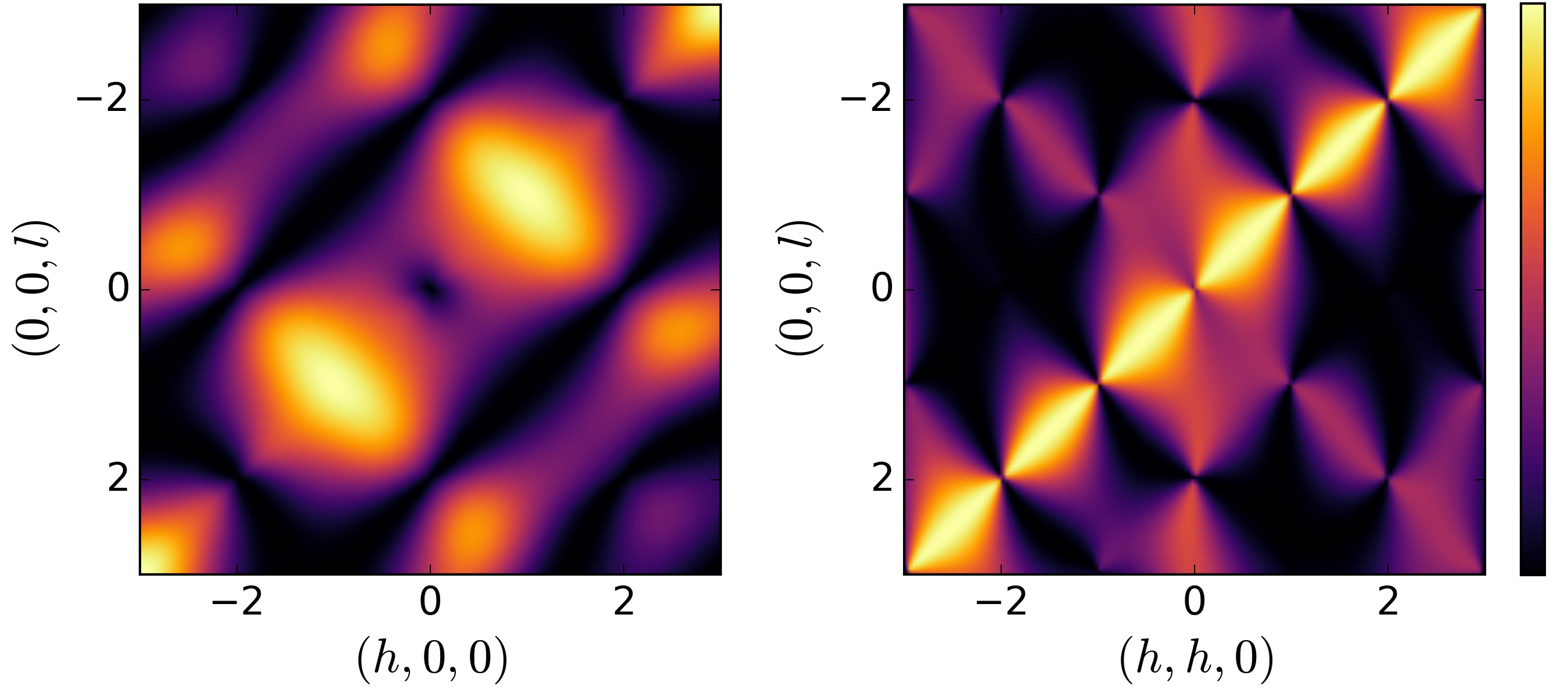}
	\caption{Equal-time spin structure factor in the spin-flip channel [Eq. (\ref{eq_structurefactors})], with an electric field in the $[111]$ direction and $\chi=1.49$. We show the $[h0l]$ plane with $\vect{\eta}_\nu = [010]$ (left), and the $[hhl]$ plane with $\vect{\eta}_\nu = [1\overline{1}0]$ (right). We use $\mathcal{W}/2Ug= 0.1$ to regularize the theory, and the rescaling parameters for the emergent magnetic field $b_m(\mathbf{k})$ are $\Lambda_1=1$ and $\Lambda_2=\Lambda_3=\Lambda_4 = 2$.}
	\label{fig_rg}
\end{figure}

With a regularized effective lattice theory at hand, we are now in a position to examine the possible experimental signatures of the external electric field near the phase transition. In Fig. \ref{fig_structurefactors}, the rightmost column shows the spin structure factors close to the critical point. Strikingly, for an external electric field in the $[111]$ direction, the {\it pinch points} re-emerge  in the $[hhl]$ plane as $\chi$ is tuned towards the critical value of $1.5$. However, no such {\it pinch points} are seen for an external field in the $[011]$ direction. As argued in Ref. \onlinecite{PhysRevB.86.075154}, we expect pinch points to appear when the photon dispersion is dominated by the quadratic behavior. This is precisely what happens here: for $\mathbf{E} \propto [111]$, the photon velocity vanishes in the $[111]$ direction for \emph{both} polarization modes, while this is not the case for $\mathbf{E} \propto [011]$ (see the $\Gamma- L$ direction in Fig. \ref{fig_lattice_reg}).

At this point, we note that our expressions for the equal-time spin structure factor (Eqs. (\ref{eq_structurefactors_0}) and (\ref{eq_structurefactors})) do not take into account the possible renormalization of the fields due to the integration of higher-energy modes. This renormalization is  accounted for by rescaling the emergent magnetic field as 
$$b_m({\bf k})\rightarrow\Lambda_m(\hat{\bf k}) b_m({\bf k}),$$
where $\Lambda_m(\hat{\bf k})$ is a rescaling factor. Since the system, even in the absence of an external electric field, does not have a full rotational symmetry (see Eq. (\ref{eq_quantum spin iceham})), the rescaling factor is in general direction dependent. However, the effective low-energy theory near ${\bf k=0}$ [Eq. (\ref{eq_cont_ham0})] has an emergent rotation symmetry and, in this limit, the rescaling factor can be chosen to be isotropic as in Refs. \onlinecite{PhysRevB.86.075154, PhysRevLett.100.047208}. In presence of an electric field, even near ${\bf k=0}$, the theory is not rotation invariant (see Eq. (\ref{eq_cont_ham})), as is evident from emergent birefringence. This renders the rescaling factor direction dependent even near ${\bf k=0}$.  Since the external electric field is even under time reversal, we must have $\Lambda_m(\hat{\bf k})=\Lambda_m(-\hat{\bf k})$. The net factor rescaling the intensity of the structure factors being quadratic in $\Lambda_m(\hat{ \mathbf{k} })$, it transforms as a biaxial nematic\cite{chaikin2000principles} with two independent parameters: one denoting the strength along the direction of the external electric field and the other perpendicular to it. However, the values of these two parameters cannot be calculated within our effective field theory, but only from more microscopic techniques such as quantum Monte Carlo.\cite{PhysRevB.86.075154, PhysRevLett.100.047208}.

 That being said, we do not expect our results to change qualitatively because of this anisotropic renormalization. Specifically, the re-emergence of the pinch points is robust under including rescaling parameters $\Lambda_m(\hat{ \mathbf{k}})$ respecting the symmetries of the biaxial nematic tensor. This is shown in Fig. \ref{fig_rg} for the particular choice $\Lambda_1 = 1$, $\Lambda_2 = \Lambda_3 = \Lambda_4 = 2$ and $\mathbf{E} \sim [111]$.


\section{Relevance to quantum spin ice Materials}
\label{sec_materials}

Are the effects described above observable? One immediate concern is that the minimal Hamiltonian in Eq. (\ref{eq_quantum spin iceham}) does not adequately describe any of the currently known quantum spin ice candidates. There are additional symmetry-allowed terms in the Hamiltonian which seem to have large coupling constants. While a calculation incorporating all these coupling constants, as well as all the allowed terms in the polarization operator [Eqs. (\ref{eq_ptrans1},\ref{eq_ptrans2},\ref{eq_ptrans3})] is beyond the scope of this work, we estimate the electric field strength required to observe the effects described within the present framework.

 In rare earth pyrochlores, typically $J_{zz}\sim 1~\mathrm{K}$ while the other couplings (including additional interactions present in real materials) are smaller. Taking $J_\pm=p~\mathrm{K}$, the strength of the quantum term [in Eqs. (\ref{eq_lattice_ham_matrixform},\ref{eq_Mmm})  in absence of the external electric field] is $\sim p^3~\mathrm{K}$.  The strength of the additional terms in the presence of the electric field is $\sim p B^2 E^2~\mathrm{K}$, where $B$ is measured in $\mathrm{(Coulomb-meter)}$ and $E$ is measured in  $\mathrm{K/(Coulomb-meter)}$. For the magnetostriction mechanism, $B\approx e u$ where $e$ is the electron charge and $u$ is the deformation of the lattice. This makes the latter term $\sim p e^2u^2 E^2~\mathrm{K}$. Thus for the two terms to be of the same order, we need
\begin{align}
E\sim\frac{p}{eu}~\mathrm{K/(Coulomb-meter)}.
\end{align}

For $u \approx 0.001-0.01\AA $ (about $0.1\%-1\%$ of a typical bond length), we get
\begin{align}
E\sim p\times (10^6-10^5)~\mathrm{V/mm} ,
\end{align}

To establish whether these effects can be observable, we note that neutron scattering experiments in the presence of an external  DC electric field of the order of $10^4~\mathrm{V/mm}$ have been reported.\cite{PhysRevB.30.1112, 0953-8984-24-43-432201} Noting that  $p<1$ in candidate systems, the effect proposed in this work seems to be in an experimentally observable realm. Regarding the effectiveness of the magnetostriction mechanism in the context of quantum spin ice materials, we note that in materials with softer triplet phonon modes, the above effect would be further enhanced. 


\section{Electric field at the surface of a non-Kramers quantum spin ice}
\label{sec_surface}

Having discussed the effect of an external electric field in the bulk of quantum spin ice, we now turn to  its potential surface effects. This is particularly striking for non-Kramers systems where the linear terms in the polarization operator (see Eqs. (\ref{eq_ptrans1}), (\ref{eq_ptrans2}) and (\ref{eq_ptrans3})) are allowed, {\it i.e.}, $L\neq 0$. The contribution of the polarization through this term is
\begin{align}
{P}_x^{(T)}=&L(-s_1^x-s_4^x+s^x_2+s^x_3)\nonumber,\\
{P}_y^{(T)}=&L\left[\frac{1}{2}(s^x_1+s^x_3-s^x_2-s^x_4)-\frac{\sqrt{3}}{2}(s_1^y+s_3^y-s_2^y-s_4^y)\right]\nonumber,\\
{P}_z^{(T)}=&L\left[\frac{1}{2}(s^x_1+s^x_2-s^x_3-s^x_4)+\frac{\sqrt{3}}{2}(s_1^y+s_2^y-s_3^y-s_4^y)\right],
\end{align}
for an up tetrahedron (and the same expression with $L\rightarrow -L$ for a down tetrahedron). Thus, for an inversion symmetric lattice without a surface,
\begin{align}
H_{\rm only~L~term}=\alpha \sum_{\boxtimes}{\bf P_{\boxtimes}^{(T)}\cdot E}=0
\end{align}
identically. However, on a surface, this is no longer the case. 

To be concrete, let us choose a [111] surface terminating in a triangular layer. Then, for the spins sitting on the last triangular layer, the contributions do not cancel and we have
\begin{align}
H_{\rm surface, L}=\alpha \sum_{i\in \triangle~surface} \bf{P}^{(T)}_i\cdot {\bf E},
\end{align}
where (in our notation):
\begin{align}
{P}_{i,x}^{(T)}&=-Ls_i^x=-\frac{L}{2}\left[s_i^++s^-_i\right],\\
{P}_{i,y}^{(T)}=&L\left[\frac{1}{2}s^x_i-\frac{\sqrt{3}}{2}s_i^y\right]=-\frac{L}{2}\left[\omega^2 s^+_i+\omega s^-_i\right],\\
{P}_{i,z}^{(T)}=&L\left[\frac{1}{2}s^x_i+\frac{\sqrt{3}}{2}s_i^y\right]=-\frac{L}{2}\left[\omega s^+_i+\omega^2s^-_i\right],
\end{align}
with $\omega=\exp[i2\pi/3]$. Due to the [111] surface termination, all surface spins are taken to belong to sublattice 1 without loss of generality. Other surfaces can be similarly considered. Let us now choose an AC electric field that acts on the surface, {\it i.e.},
\begin{align}
{\bf E}=E_0({\bf r})\cos(\Omega t) \hat{\mathbf{n}},
\end{align}
where $\hat{\bf n}$ is the direction of the electric field, and the field profile $E_0(\mathbf{r})$ is chosen appropriately to decay away from the surface into the bulk. 
	
It is clear that the linear terms create monopoles and antimonopoles on the tetrahedra to which the surface spins belong. The frequency of the spin flips is equal to $\Omega$. This corresponds to an oscillation in surface monopole density, which then results in emergent radiation that can propagate inside the bulk of the system. Thus the above protocol achieves the conversion of an AC signal of real electric fields to that of emergent fields. 

However, in candidate materials, the situation is expected to be more complicated. Apart from the fact that additional interactions need to be considered, surface imperfections and the possibility of polar surfaces also need to be  accounted for. However, given the rapid development in material sciences and the striking nature of the above effect, we hope that the present discussion will motivate future exploration in this direction.

\section{Summary and outlook}
\label{sec_summary}

In this paper, we investigated various effects of applying an external electric field on the minimal quantum spin ice Hamiltonian, exhibiting a $U(1)$ QSL ground state, which could be realized in some rare-earth pyrochlores. Using a symmetry-based approach, we obtained the general form of the effective low-energy electric polarization operator in these systems, as a function of the spin operators. Such electric polarization then couples to an external electric field, leading to qualitatively new effects in the quantum spin ice ground state, as well as in the low-energy excitations -- the gapless emergent photons.

For small electric fields, the speed of the emergent photons becomes both polarization as well as direction dependent, in striking analogy with birefringent materials. On increasing the external electric field further, the photon velocity vanishes, signaling an instability of the low-field, zero-flux QSL state. The high-field states trap $\pi$-electric fluxes on certain plaquettes, the detailed pattern of these fluxes depending on the direction of the external electric field. The high-field states are thus new types of time-reversal invariant three-dimensional $U(1)$ QSLs (with gapped electric and magnetic charges)  which are different from the low-field state in terms of the projective symmetry classification of QSLs. 

The quantum phase transition between the low-field and high-field QSLs is associated with the re-emergence of a subset of the pinch points characteristic of classical spin ice. Single electric and magnetic charges remain gapped throughout the transition. This represents a (most likely discontinuous) transition between different long-range entangled phases of condensed matter. In addition, we found that the typical field magnitudes required to observe the effects studied here are in an experimentally accessible regime. 

In this work, we elucidated the novel effects of an external electric field on quantum spin ice systems within a minimal approach.  The Hamiltonian describing candidate materials such as Yb$_2$Ti$_2$O$_7$ \cite{applegate2012vindication,PhysRevB.70.180404,ross2011quantum,chang2012higgs,thompson2011rods} and Tb$_2$Ti$_2$O$_7$\cite{fennell2012power,PhysRevB.86.174403,legl2012vibrating,PhysRevLett.82.1012,PhysRevB.68.180401}  likely include terms beyond the ones considered here, the effect of which  may be worth analyzing in detail. However,  the fact that new candidate materials are constantly being added -- such as most recently Pr$_2$Hf$_2$O$_7$\cite{PhysRevB.94.024436, PhysRevB.94.144415, QSIobservation} -- is cause for optimism  that the physics discussed here will be subjected to experimental scrutiny in the not too distant future. 

In the same spirit, it is useful to note that in this work, a term linear in the external electric field does not occur in the effective low-energy Hamiltonian [Eq. (\ref{eq_lattice_ham})] because the underlying system/phase does not break inversion symmetry explicitly/spontaneously and hence cannot have a net electric dipole moment. This is a crucial difference with the case of an applied \textit{magnetic} field, which normally couples linearly to the underlying pseudo-spin degrees of freedom due to explicit breaking of time-reversal symmetry. Such Zeeman coupling, although in principle can lead to a magnetic analog of the anisotropic responses studied here, usually has a natural instability to magnetically polarized phases and leads to qualitatively different effects than phase transitions between QSLs \cite{QSImagneticfield}. (The coupling to electromagnetic \textit{radiation} was also explored recently in the context of Raman scattering\cite{PhysRevB.96.035136}.)  

However, inversion symmetry is explicitly broken in a class of pyrochlore magnets dubbed {\it breathing pyrochlores} ({\it e.g.}, LiGaCr$_4$O$_8$\cite{PhysRevLett.110.097203,PhysRevLett.113.227204}, LiInCr$_4$O$_8$\cite{PhysRevLett.110.097203,PhysRevLett.113.227204}, and Ba$_3$Yb$_2$Zn$_5$O$_{11}$\cite{PhysRevLett.116.257204}), where the effect of such linear terms may lead to interesting experimental consequences, particularly in the context of quantum spin liquid physics.\cite{PhysRevB.94.075146}

\section*{Acknowledgments}
We thank K. Damle, L. Jaubert, I. Mandal and P.V. Sriluckshmy for useful discussions.
\'{E}. L.-H. acknowledges funding from FRQNT, NSERC and support from the Stewart Blusson Quantum Matter Institute. S. B. and \'{E}. L.-H. are grateful for the hospitality of the visitors program at MPIPKS, funded in part  via the Leibniz Prize program of the Deutsche Forschungsgemeinschaft,  where most of this work was completed. R. M. was in part supported by the Deutsche Forschungsgemeinschaft via SFB 1143 and S. B. was in part supported by the start-up grant at ICTS-TIFR.


\appendix


\section{The pyrochlore lattice}
\label{appen_pyrochlorelattice}

\begin{figure}[h]
	\centering
	\subfigure[]{
		\includegraphics[width=0.43\columnwidth]{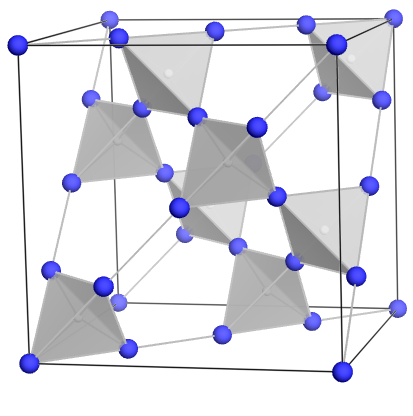}
		\label{fig_lattice_a}
	}
	\subfigure[]{
		\includegraphics[width=0.5\columnwidth]{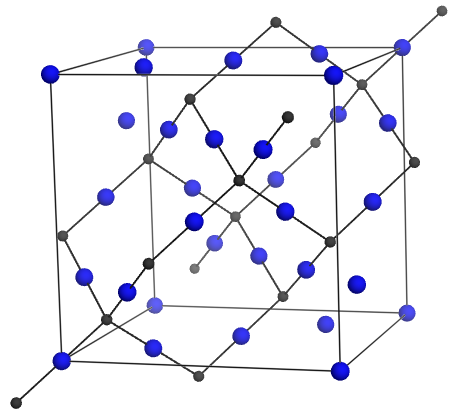}
		\label{fig_lattice_b}
	}
	\caption{The pyrochlore lattice, whose sites (where spin ice magnetic moments are located)  are depicted in blue. One cubic unit cell of side $a_0$ is shown. (a): The pyrochlore lattice realizes a network of corner-sharing tetrahedra, shown in gray. (b): The centers of those tetrahedra form the \textit{medial} diamond lattice (shown in black). The magnetic moments sit on the bonds of this diamond lattice, and the local spin quantization axes point in the direction of the bond.}
	\label{fig_lattice}
\end{figure}

The magnetic moments of (quantum) spin ice are located on the sites of the pyrochlore lattice, a network of corner-sharing tetrahedra with alternating `up' and `down' orientation, as shown in Fig. \ref{fig_lattice_a}. In other words, the pyrochlore lattice is a FCC lattice, but decorated at each site with an `up' (say) tetrahedron.
The primitive unit cell is a single tetrahedron with four sublattice sites, as shown in Fig. \ref{fig_axes}. The local spin quantization axes are different for each sublattice site; the explicit expressions are given below. It is also useful to define a cubic unit cell, of side $a_0$, that comprises 16 lattice sites.

The centers of the tetrahedra form a diamond lattice (referred to as the \textit{medial} or \textit{direct} diamond lattice in the text), as shown in Fig. \ref{fig_lattice_b}. The magnetic moments (or pyrochlore lattice sites) are located on the middle of the diamond lattice bonds, whereas the local spin quantization axes point in the diamond bond directions (that is, `in' or `out' of every tetrahedron).

\subsection{Local spin quantization axes}
\label{appen_latticedetails}

\begin{figure}[h]
	\centering
	\includegraphics[scale=0.3]{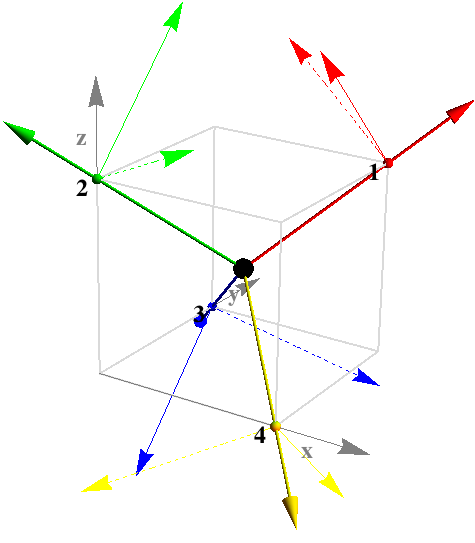}
	\caption{The local spin quantization axes for spin ice. Thick lines denote $\hat{\bf t}_i$, the thin lines denote $\hat{\bf y}_i$ and the dashed lines denote $\hat{\bf x}_i$. These directions are given by Eqs. \ref{eq_t_axes} and \ref{eq_xy_axes}}
	\label{fig_axes}
\end{figure}
The local quantization axes, $\hat{\bf t}_i$ (as shown in Fig. \ref{fig_axes}), are given  by 
\begin{align}
& \hat{\bf t}_1=\frac{1}{\sqrt{3}}[111], \quad \hat{\bf t}_2=\frac{1}{\sqrt{3}}[\bar1\bar11], \nonumber \\
&\hat{\bf t}_3=\frac{1}{\sqrt{3}}[\bar11\bar1], \quad \hat{\bf t}_4=\frac{1}{\sqrt{3}}[1\bar1\bar1].
\label{eq_t_axes}
\end{align}

Therefore, the local axes $\hat{\bf x}_i$ and $\bf{\hat y}_i$ that form the local triads are given by (as shown in Fig. \ref{fig_axes})
\begin{align}
&\hat{\bf x}_1=\frac{1}{\sqrt{6}}[\bar211], \quad \hat{\bf y}_1=\frac{1}{\sqrt{2}}[0\bar11],\nonumber\\
&\hat{\bf x}_2=\frac{1}{\sqrt{6}}[2\bar11], \quad \hat{\bf y}_2=\frac{1}{\sqrt{2}}[011],\nonumber\\
&\hat{\bf x}_3=\frac{1}{\sqrt{6}}[21\bar1], \quad \hat{\bf y}_3=\frac{1}{\sqrt{2}}[0\bar1\bar1],\nonumber\\
&\hat{\bf x}_4=\frac{1}{\sqrt{6}}[\bar2\bar1\bar1], \quad \hat{\bf y}_4=\frac{1}{\sqrt{2}}[01\bar1].
\label{eq_xy_axes}
\end{align}

\section{The polarization operator}
\label{appen_polar}

There are 24 elements in the tetrahedral group, $\mathcal{T}_d$, divided into 5 classes. These are
\begin{align}
\mathcal{T}_d:E;\{c_3,c_3^2\}(8);\{S_4,S_4^3\}(6);\{S_4^2=C_2\}(3);\{\sigma_d\}(6) .
\end{align}
Hence there are 5 irreducible representations: $A_1,A_2,E,T_1,T_2$. The octahedral group is obtained by taking the direct  product of $\mathcal{T}_d$ with the inversion group $I$, {\it i.e.}, $O_h\simeq\mathcal{T}_d\bigotimes I$. Thus the representations of $\mathcal{T}_d$ can be further classified according to  their behavior, {\it i.e.}, even ($g$) or odd ($u$), under inversion. 

Consider a single up tetrahedron. The transformation of the $s^z$ component under various generators of the point group symmetries of $\mathcal{T}_d$ are given by

\begin{align}
C_3[111]&: \{s^z_1,s^z_2,s^z_3,s^z_4\}\rightarrow\{s^z_1,s^z_4,s^z_2,s^z_3\}\nonumber,\\
C_2[\hat{\bf z}]&:\{s^z_1,s^z_2,s^z_3,s^z_4\}\rightarrow\{s^z_2,s^z_1,s^z_4,s^z_3\},\\
S_4[-\hat{\bf z}]&:\{s^z_1,s^z_2,s^z_3,s^z_4\}\rightarrow\{s^z_3,s^z_4,s^z_2,s^z_1\}\nonumber,\\
\sigma_d[x=y]&:\{s^z_1,s^z_2,s^z_3,s^z_4\}\rightarrow\{s^z_1,s^z_2,s^z_4,s^z_3\}\nonumber,
\end{align}
while the transverse components transform as
\begin{widetext}
\begin{align}
C_3[111]&:~~~~\left\{\begin{array}{c}
s_1^x\rightarrow-\frac{1}{2}s_1^x+\frac{\sqrt{3}}{2}s_1^y;~~~~~s_1^y\rightarrow-\frac{\sqrt{3}}{2}s_1^x-\frac{1}{2}s_1^y\\
s_2^x\rightarrow-\frac{1}{2}s_4^x+\frac{\sqrt{3}}{2}s_4^y;~~~~~s_2^y\rightarrow-\frac{\sqrt{3}}{2}s_4^x-\frac{1}{2}s_4^y\\
s_3^x\rightarrow-\frac{1}{2}s_2^x+\frac{\sqrt{3}}{2}s_2^y;~~~~~s_3^y\rightarrow-\frac{\sqrt{3}}{2}s_2^x-\frac{1}{2}s_2^y\\
s_4^x\rightarrow-\frac{1}{2}s_3^x+\frac{\sqrt{3}}{2}s_3^y;~~~~~s_4^y\rightarrow-\frac{\sqrt{3}}{2}s_3^x-\frac{1}{2}s_3^y\\
{\rm or}\\
s_1^{\pm}\rightarrow e^{\mp i\frac{2\pi}{3}}s_1^\pm;~~~~s_2^{\pm}\rightarrow e^{\mp i\frac{2\pi}{3}}s_4^\pm;~~~~s_3^{\pm}\rightarrow e^{\mp i\frac{2\pi}{3}}s_2^\pm;~~~~s_4^{\pm}\rightarrow e^{\mp i\frac{2\pi}{3}}s_3^\pm
\end{array}\right.
\\
C_2({\bf z})&:~~~~\left\{\begin{array}{c}
s_1^x\rightarrow s_2^x;~~~~s_1^y\rightarrow s_2^y\\
s_2^x\rightarrow s_1^x;~~~~s_2^y\rightarrow s_1^y\\
s_3^x\rightarrow s_4^x;~~~~s_3^y\rightarrow s_4^y\\
s_4^x\rightarrow s_3^x;~~~~s_4^y\rightarrow s_3^y\\
{\rm or}\\
s_1^\pm\rightarrow s_2^\pm;~~~s_2^\pm\rightarrow s_1^\pm;~~~s_3^\pm\rightarrow s_4^\pm;~~~s_4^\pm\rightarrow s_3^\pm
\end{array}\right.
\\
S_4[-\hat{\bf z}]&:~~~~\left\{\begin{array}{c}
s_1^x\rightarrow -\frac{1}{2}s_3^x+\frac{\sqrt{3}}{2}s^y_3;~~~~s^y_1\rightarrow\frac{\sqrt{3}}{2}s_3^x+\frac{1}{2}s_3^y\\
s_2^x\rightarrow -\frac{1}{2}s_4^x+\frac{\sqrt{3}}{2}s^y_4;~~~~s^y_2\rightarrow\frac{\sqrt{3}}{2}s_4^x+\frac{1}{2}s_4^y\\
s_3^x\rightarrow -\frac{1}{2}s_2^x+\frac{\sqrt{3}}{2}s^y_2;~~~~s^y_3\rightarrow\frac{\sqrt{3}}{2}s_2^x+\frac{1}{2}s_2^y\\
s_4^x\rightarrow -\frac{1}{2}s_1^x+\frac{\sqrt{3}}{2}s^y_1;~~~~s^y_4\rightarrow\frac{\sqrt{3}}{2}s_1^x+\frac{1}{2}s_1^y\\
{\rm or}\\
s_1^\pm\rightarrow e^{\pm i 2\pi/3}s_3^\mp;~~~s_2^\pm\rightarrow e^{\pm i 2\pi/3}s_4^\mp;~~~s_3^\pm\rightarrow e^{\pm i 2\pi/3}s_2^\mp;~~~s_4^\pm\rightarrow e^{\pm i 2\pi/3}s_1^\mp
\end{array}\right.
\\
\sigma_d[x=y]&:~~~~\left\{\begin{array}{c}
s_1^x\rightarrow -\frac{1}{2}s_1^x+\frac{\sqrt{3}}{2}s^y_1;~~~~s^y_1\rightarrow\frac{\sqrt{3}}{2}s_1^x+\frac{1}{2}s_1^y\\
s_2^x\rightarrow -\frac{1}{2}s_2^x+\frac{\sqrt{3}}{2}s^y_2;~~~~s^y_2\rightarrow\frac{\sqrt{3}}{2}s_2^x+\frac{1}{2}s_2^y\\
s_3^x\rightarrow -\frac{1}{2}s_4^x+\frac{\sqrt{3}}{2}s^y_4;~~~~s^y_3\rightarrow\frac{\sqrt{3}}{2}s_4^x+\frac{1}{2}s_4^y\\
s_4^x\rightarrow -\frac{1}{2}s_3^x+\frac{\sqrt{3}}{2}s^y_3;~~~~s^y_4\rightarrow\frac{\sqrt{3}}{2}s_3^x+\frac{1}{2}s_3^y\\
{\rm or}\\
s_1^\pm\rightarrow e^{\pm i 2\pi/3}s_1^\mp;~~~s_2^\pm\rightarrow e^{\pm i 2\pi/3}s_2^\mp;~~~s_3^\pm\rightarrow e^{\pm i 2\pi/3}s_4^\mp;~~~s_4^\pm\rightarrow e^{\pm i 2\pi/3}s_3^\mp.
\end{array}\right.
\end{align}
\end{widetext}

These give two contributions to the polarization operator, which transforms under the $T_{1u}$ representation of the octahedral group. We can then write the polarization operator for the up tetrahedron as given in Eq. (\ref{eq_polop}) where the different contributions are given by Eqs. (\ref{eq_plong}), (\ref{eq_ptrans1}), (\ref{eq_ptrans2}), and (\ref{eq_ptrans3}).

\section{Perturbation theory in quantum spin ice with a uniform electric field}
\label{appen_pert1}

\begin{figure}
\centering
\includegraphics[scale=0.42]{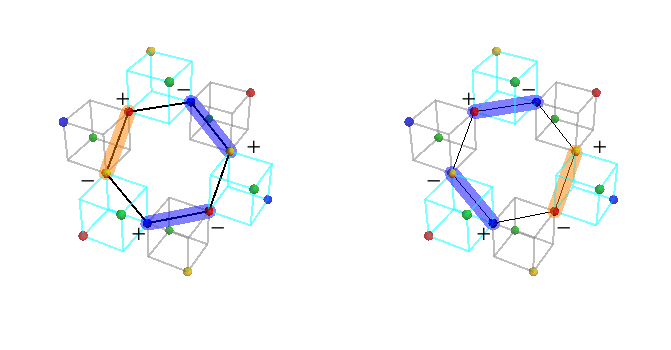}
\caption{Two typical terms of order $\alpha J_\pm^2$ coming from the same hexagon. The $\pm$ labels denote the contribution of $s^\pm$ at each site. The blue bonds denote the contributions coming from $H_\pm$ while the orange bonds denote the contributions coming from $H_{\bf E}$. Because of inversion symmetry, the mutual coefficients of the orange bonds have a relative minus sign since one of them comes from an up tetrahedron (in gray) while the other comes from a down tetrahedron (in cyan). Therefore they cancel pairwise.}
\label{fig_lin3}
\end{figure} 

\begin{figure}
	\centering
	\subfigure[]{
		\includegraphics[scale=0.42]{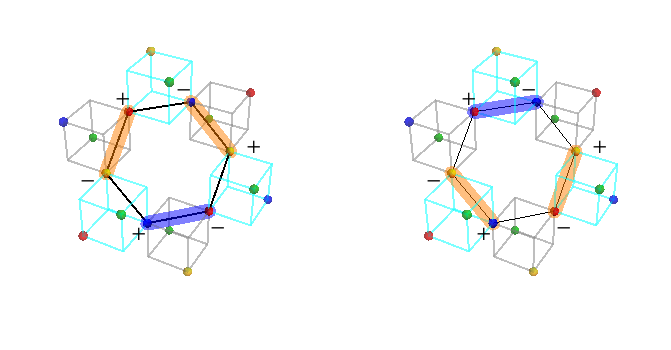}
	}
	\subfigure[]{
		\includegraphics[scale=0.42]{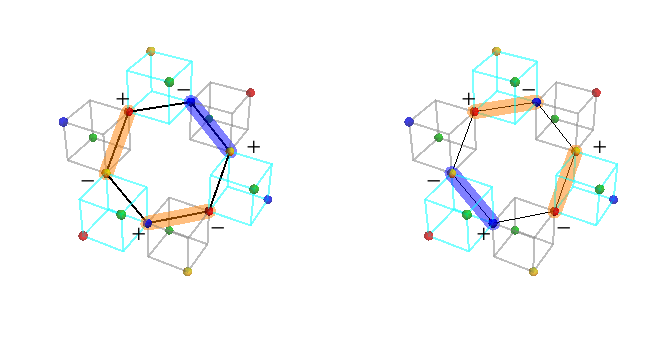}
	}
	\subfigure[]{
		\includegraphics[scale=0.42]{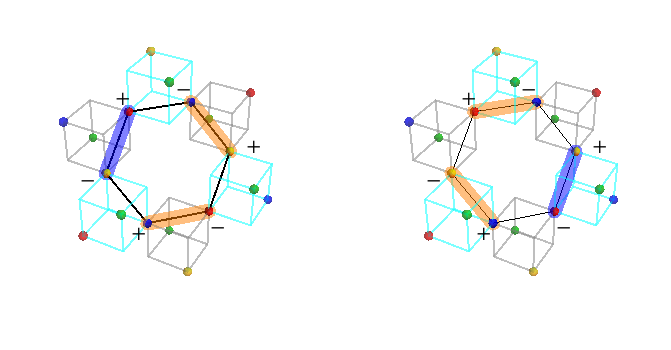}
	}
	\caption{Summing all the contributions of order $\alpha^2 J_\pm$ coming from the hexagons perpendicular to $\hat{\bf t}_3$. Because the hexagons now have \emph{two} orange bonds, the contributions from the left and right columns \emph{add} up, for each line (a), (b) and (c). The total contribution is therefore given by  The other 3 three types of hexagons, perpendicular to  $\hat{\bf t}_1$,  $\hat{\bf t}_2$ and  $\hat{\bf t}_4$, work similarly.}
	\label{fig_qu3}
\end{figure}

In the limit where $J_{zz}$ is larger than $J_\pm$ and $|{\bf E}|$ in the Hamiltonian in Eq. (\ref{eq_haminfield}), we can derive the effective low-energy theory (below the energy scale of $O(J_{zz})$). This effective Hamiltonian is given by
\begin{align}
\mathcal{H}_{\text{eff}}=\mathcal{P}\left[H_{p}+H_pG_0'H_p+\cdots\right]\mathcal{P},
\end{align}
where $\mathcal{P}$ is the projector to the ice manifold and 
\begin{align}
H_p&=H_{\bf E} - J_\pm\sum_{\left< ij\right>}(s^+_is^-_j+s^-_is^+_j),\nonumber\\
G_0'&=(1-\mathcal{P})\frac{1}{E-H_{zz}}(1-\mathcal{P}),\nonumber\\
H_{zz}&= J_{zz}\sum_{\left< ij\right>} s^z_is^z_j.
\end{align}

\paragraph{\underline{First order} :}
At the first order, all terms in $H_p$ take the state out of the ice manifold and ${\bf P}^{(L)}$ is identically zero in this manifold. So there are no first order contributions.

\paragraph{\underline{Second order} :}
The second order term has the form
\begin{align}
\mathcal{H}_{\text{eff}}^{(2)}=\mathcal{P}\left[H_pG_0'H_p\right]\mathcal{P}.
\end{align}

There are three types of contributions:
\begin{enumerate}
\item $O(J_\pm^2)$: This term is a constant.\cite{PhysRevB.69.064404}
\item $O(\alpha J_\pm)$: To calculate this term, it is enough to consider an electric field along the z-direction, {\it i.e.}, ${\bf E}=E\hat{\bf z}$. We immediately notice that the longitudinal part cannot contribute since $P^{(L)}_z\mathcal{P}=0$ identically. The transverse terms cancel.

\item $O(\alpha^2)$: These lead to contributions of the form $\sim{\bf E\cdot E}$, which renormalize the dielectric constants. Such terms generally arise at all even orders and hence they will be not discussed here, though they can change the dielectric properties of the system in response to DC electric fields.
\end{enumerate}

\paragraph{Third order}
The third order term has the form
\begin{align}
\mathcal{H}_{\text{eff}}^{(3)}=\mathcal{P}\left[H_pG_0'H_pG_0'H_p\right]\mathcal{P}.
\end{align}

The non-trivial contributions come from the hexagons.  The 4 types of contributions are:
\begin{enumerate}
\item $O(J_\pm^3)$: This is the usual ring-exchange term\cite{PhysRevB.69.064404},
\begin{align}
-12\frac{J_\pm^3}{J_{zz}^2}\sum_{{\hexagon},(1,\dots 6\in\hexagon)}\left(s_1^+s_2^-s_3^+s_4^-s_5^+s_6^-+\text{h.c.}\right).
\end{align}

\item $O(\alpha J^2_\pm)$: The terms at this order cancel out (as shown in Fig. \ref{fig_lin3}) due to inversion symmetry.

\item $O(\alpha^3)$: Due to inversion symmetry, these terms too cancel pairwise.

\item $O(\alpha^2 J_\pm)$: The contribution from this term is (as shown in Fig. \ref{fig_qu3}):

\begin{align}
&- 6 \frac{\alpha^2J_\pm B^2 \left[3({\bf E}\cdot\hat{\bf t}_1)^2-{\bf E\cdot E}\right]}{J_{zz}^2}\sum_{{\hexagon}\perp\hat{\bf t}_1}(\hat{\mathcal{O}}_{\hexagon}+\text{h.c.})\nonumber\\
& - 6 \frac{\alpha^2J_\pm B^2 \left[3({\bf E}\cdot\hat{\bf t}_2)^2-{\bf E\cdot E}\right]}{J_{zz}^2}\sum_{{\hexagon}\perp\hat{\bf t}_2}(\hat{\mathcal{O}}_{\hexagon}+\text{h.c.})\nonumber\\
& - 6 \frac{\alpha^2J_\pm B^2 \left[3({\bf E}\cdot\hat{\bf t}_3)^2-{\bf E\cdot E}\right]}{J_{zz}^2}\sum_{{\hexagon}\perp\hat{\bf t}_3}(\hat{\mathcal{O}}_{\hexagon}+\text{h.c.}) \nonumber\\
& - 6 \frac{\alpha^2J_\pm B^2 \left[3({\bf E}\cdot\hat{\bf t}_4)^2-{\bf E\cdot E}\right]}{J_{zz}^2}\sum_{{\hexagon}\perp\hat{\bf t}_4}(\hat{\mathcal{O}}_{\hexagon}+\text{h.c.}),
\end{align}
where $\hat{\mathcal{O}}_{\hexagon} = s_1^+s_2^-s_3^+s_4^-s_5^+s_6^ , (1,\dots 6\in\hexagon)$ and the four contributions come from four types of hexagons that are perpendicular to $\hat{\bf t}_i$ for $i=1,2,3,4$.
\end{enumerate}

\section{Emergent photons under a uniform electric field}

\subsection{Continuum theory}
\label{appen_continuum_photons}
Here we derive the photon dispersion in the continuum limit for the Hamiltonian given by Eq. (\ref{eq_cont_ham}). Using Eqs. (\ref{eq_continuum_e}) and (\ref{eq_continuum_b}) for the emergent fields $e$ and $b$ respectively, the electric and magnetic contributions to the Hamiltonian become:
\begin{widetext}
	\begin{align}
	\mathcal{H}_{e^2} &= \frac{1}{4}  \sum_{\lambda, \lambda'} \int \frac{d^3\mathbf{k}}{(2\pi)^3} \sqrt{\omega_\lambda(\mathbf{k})\omega_{\lambda'}(\mathbf{k})} \sum_\sigma \left[ \left[{\epsilon}^\sigma_{\lambda'}(\mathbf{k})\right]^* {\epsilon}^\sigma_\lambda(\mathbf{k})  a_{\lambda \mathbf{k}} a_{\lambda' \mathbf{k}}^\dagger - {\epsilon}^\sigma_{\lambda'} (-\mathbf{k}) \epsilon_\lambda^\sigma(\mathbf{k})  a_{\lambda \mathbf{k}} a_{\lambda', -\mathbf{k}}  + \text{h.c.} \right],
	\label{eq_app_cont_e2}
	\end{align}
\begin{align}
\mathcal{H}_{b^2}&= \frac{1}{4} \sum_{\lambda, \lambda'} \int \frac{d^3\mathbf{k}}{(2\pi)^3} \frac{ \mathbf{k}^2}{\sqrt{\omega_\lambda(\mathbf{k})\omega_{\lambda'}(\mathbf{k})}} \sum_{\sigma,\rho} Q_{\sigma \rho}(\hat{\mathbf{k}}) \left[ \left[ {\epsilon}^\sigma_{\lambda'}(\mathbf{k})\right]^* {\epsilon}^\rho_\lambda(\mathbf{k})  a_{\lambda \mathbf{k}} a_{\lambda' \mathbf{k}}^\dagger 
+ {\epsilon}^\sigma_{\lambda'} (-\mathbf{k}) \epsilon_\lambda^\rho(\mathbf{k})  a_{\lambda \mathbf{k}} a_{\lambda', -\mathbf{k}}  +  \text{h.c.} \right],
	\label{eq_app_cont_b2}
\end{align}
\end{widetext}
where we defined $Q(\hat{\mathbf{k}}) = \mathcal{U} \Xi^T R(\hat{\mathbf{k}}) \Xi$, which is a Hermitian matrix, with
\begin{align} 
R(\hat{\mathbf{k}}) = \begin{pmatrix}
1 - \hat{k}_x^2     & -\hat{k}_x\hat{k}_y & -\hat{k}_x\hat{k}_z \\
-\hat{k}_x\hat{k}_y & 1 - \hat{k}_y^2     & -\hat{k}_y\hat{k}_z \\
-\hat{k}_x\hat{k}_z & -\hat{k}_y\hat{k}_z & 1 - \hat{k}_z^2
\end{pmatrix},
\end{align}
and $\hat{\mathbf{k}} \equiv \mathbf{k}/|\mathbf{k}|$.

We can now choose a specific basis for the photon polarization vectors $ \vect{\epsilon}_\lambda(\mathbf{k})$. We remark that choosing such a basis amounts to choosing an `electromagnetic' gauge because, using Eq. (\ref{eq_continuum_A}), the divergence of the vector potential $\mathbf{A}$ is now fixed. For convenience, we choose the polarization vectors to be the eigenvectors of $Q(\hat{\mathbf{k}})$ with positive eigenvalues $\xi^2_\lambda(\hat{\mathbf{k}})$ -- this is allowed because $Q$ is a positive-definite, Hermitian matrix (for small applied electric fields). Then:
\begin{align}
\sum_\rho Q_{\sigma \rho}(\hat{\mathbf{k}}) \epsilon^\rho_\lambda(\mathbf{\hat{k}}) =  \xi^2_\lambda(\mathbf{\hat{k}}) \epsilon^\sigma_\lambda(\mathbf{\hat{k}}), 
\end{align} 
and, because $Q(-\hat{\mathbf{k}})$ = $Q(\hat{\mathbf{k}})^*$, its eigenvectors satisfy $$  \epsilon^\sigma_\lambda(-\mathbf{\hat{k}}) = \left[ \epsilon^\sigma_\lambda(\mathbf{\hat{k}}) \right]^*,$$ 
while the unitarity of the matrix of eigenvectors enforces
$$ \sum_\sigma \left[\epsilon^\sigma_{\lambda'}(\hat{\mathbf{k}})\right]^* \epsilon^\sigma_\lambda(\hat{\mathbf{k}})  = \delta_{\lambda \lambda'} . $$

Using the above identities and putting together the electric and magnetic contributions, Eqs. (\ref{eq_app_cont_e2}) and (\ref{eq_app_cont_b2}) respectively, we get:
\begin{align}
\mathcal{H}_\text{eff}^\text{cont} &= \frac{1}{4} \sum_\lambda \int \frac{d^3\mathbf{k}}{(2\pi)^3} \bigg[ \left( \frac{\mathbf{k}^2}{\omega_\lambda(\mathbf{k})} \xi^2_\lambda(\hat{\mathbf{k}}) + \omega_\lambda(\mathbf{k}) \right) a_{\lambda \mathbf{k}}  a_{\lambda \mathbf{k}}^\dagger \nonumber \\ &+ \left( \frac{ \mathbf{k}^2}{\omega_\lambda(\mathbf{k})} \xi^2_\lambda(\hat{\mathbf{k}}) - \omega_\lambda(\mathbf{k}) \right)  a_{\lambda \mathbf{k}}  a_{\lambda, -\mathbf{k}} +  \text{h.c.} \bigg].
\end{align}
Finally, the photon dispersion relation is obtained by requiring that the terms which do not conserve photon number vanish, {\it i.e.},
\begin{align}
	\omega_\lambda(\mathbf{k}) =  |\xi_\lambda(\hat{\mathbf{k}})| |\mathbf{k}|,
\end{align}
which is just Eq. (\ref{eq_dispersion_continuum}) in the main text. The Hamiltonian then assumes the usual form,
\begin{align}
\mathcal{H}_\text{eff}^\text{cont} &=\sum_\lambda \int \frac{d^3\mathbf{k}}{(2\pi)^3}  \omega_\lambda(\mathbf{k}) \Big[a_{\lambda \mathbf{k}}^\dagger a_{\lambda \mathbf{k}} + \frac{1}{2} \Big].
\end{align}

\subsection{Lattice theory}
\label{appen_lattice_photons}
\subsubsection{Computation of the lattice magnetic field}
On the pyrochlore lattice, the magnetic field and the vector potential are related by Eq. (\ref{eq_lattice_ba}): 
$$b_{(\mathbf{r},n)} =(\nabla_{\hexagon} \times A_{(\mathbf{s},m)})_{(\mathbf{r},n)}. $$

To compute this curl, we take the sum of the fields $A_{(\mathbf{s},m)}$ living on the six bonds of an hexagonal plaquette. These bonds have midpoints -- corresponding to sites on the dual diamond lattice spanned by $(\mathbf{s},m)$ -- located at
\begin{align}
 \left( \mathbf{r} - \frac{\mathbf{t}_n}{2} \right) \pm \mathbf{h}_{nm}  \quad , \quad \mathbf{h}_{nm} = \frac{a_0}{\sqrt{8}} \frac{\hat{\mathbf{t}}_n \times \hat{\mathbf{t}}_m}{|\hat{\mathbf{t}}_n \times \hat{\mathbf{t}}_m |},
 \end{align}
where $\left( \mathbf{r} - \mathbf{t}_n/2 \right)$ represents the middle of the hexagonal plaquette, $\mathbf{h}_{nn} = 0$ and $\mathbf{h}_{nm} = -\mathbf{h}_{mn}$ by construction. Here, the $\pm$ encodes the fact that the 6 sites forming an hexagonal plaquette are arranged by pairs, at the same distance but in opposite directions from the middle of the hexagon. When summing over the 6 plaquette bonds, the sign of neighboring terms alternate because they come either from an up $\rightarrow $ down bond or a down $\rightarrow$ up bond (remember that $A_{(\mathbf{s},m)}$ is a directed variable). Using Eq. (\ref{eq_lattice_a}) for the gauge potential, we get:
\begin{widetext}
	\begin{align}
	b_{(\mathbf{r},n)}
	&= \sqrt{\frac{2}{N}} \sum_{\mathbf{k}, \lambda, m}  \sqrt{\frac{\mathcal{M}_m}{\omega_\lambda(\mathbf{k})}} \Big[ e^{-i\mathbf{k} \cdot (\mathbf{r} - \mathbf{t}_n/2)} \big\{ -2i \sin(\mathbf{k} \cdot \mathbf{h}_{nm}) \big\} \eta_{m \lambda}(\mathbf{k}) a_{\lambda \mathbf{k}} + \text{h.c.} \Big].
	\end{align}

This expression is simplified by introducing a Hermitian, anti-symmetric matrix\cite{PhysRevB.86.075154}:
\begin{align}
Z(\mathbf{k}) = -2i \begin{pmatrix} 0 & s_{01}(\mathbf{k}) & s_{02}(\mathbf{k}) & s_{03}(\mathbf{k}) \\
-s_{01}(\mathbf{k}) & 0 & s_{12}(\mathbf{k}) & s_{13}(\mathbf{k}) \\
-s_{02}(\mathbf{k}) & -s_{12}(\mathbf{k}) & 0 & s_{23}(\mathbf{k}) \\
-s_{03}(\mathbf{k}) & -s_{13}(\mathbf{k}) & -s_{23}(\mathbf{k}) & 0 \end{pmatrix},
\label{eq_Zmatrix_definition}
\end{align}
with $s_{nm}(\mathbf{k}) \equiv \sin(\mathbf{k} \cdot \mathbf{h}_{nm})$. Using this, we get Eq. (\ref{eq_lattice_b}).

%
\subsubsection{Photon Dispersion}
Here we review how to derive the dispersion relation for lattice photons [Eq. (\ref{eq_lattice_dispersion})], starting from the lattice Hamiltonian for quantum spin ice [Eq. (\ref{eq_lattice_ham_matrixform_expanded})]. Using Eqs. (\ref{eq_lattice_e}) and (\ref{eq_lattice_b}) for the emergent electric and magnetic fields, respectively, we calculate the electric and magnetic contributions to the Hamiltonian (note that the sums over $\mathbf{r}$ and $\mathbf{s}$ in Eq. (\ref{eq_lattice_ham_matrixform_expanded}) run over $N/4$ up tetrahedra):
\begin{align}
\mathcal{H}_{e^2} &= \frac{1}{4} \sum_{\mathbf{k}, \lambda, \lambda'} \sqrt{\omega_\lambda(\mathbf{k}) \omega_{\lambda '}(\mathbf{k})} \sum_m \left[ \eta^*_{m \lambda'}(\mathbf{k})  \eta_{m \lambda} (\mathbf{k})  a_{\lambda \mathbf{k}} a_{\lambda' \mathbf{k}}^\dagger -\eta_{m \lambda'}(-\mathbf{k}) \eta_{m \lambda} (\mathbf{k})  a_{\lambda \mathbf{k}}  a_{\lambda', -\mathbf{k}} + \text{h.c.} \right],
\label{eq_app_lattice_e2}
\end{align}
	\begin{align}
	\mathcal{H}_{b^2}  & = \frac{1}{4}  \sum_{\mathbf{k}, \lambda, \lambda '} \frac{1}{ \sqrt{\omega_\lambda(\mathbf{k}) \omega_{\lambda'}(\mathbf{k})} } \sum_{m',m} T_{m'm}(\mathbf{k}) \left[ \eta^*_{m' \lambda'}(\mathbf{k})  \eta_{m \lambda}(\mathbf{k})  a_{\lambda \mathbf{k}}  a_{\lambda' \mathbf{k}}^\dagger + \eta_{m' \lambda'}(-\mathbf{k}) \eta_{m \lambda}(\mathbf{k})  a_{\lambda \mathbf{k}} a_{\lambda' ,-\mathbf{k}} + \text{h.c.}  \right].
	\label{eq_app_lattice_b2}
	\end{align}
\end{widetext}
For simplicity of notation, we defined a Hermitian matrix $T(\mathbf{k})$ whose elements are given by
\begin{align}
 T_{m'm}(\mathbf{k}) = U \sum_n \sqrt{\mathcal{M}_{m'}} Z_{m'n}(\mathbf{k}) Z_{nm}(\mathbf{k}) \sqrt{\mathcal{M}_m},
 \end{align}
and used the property $Z_{nm'}(-\mathbf{k}) = Z_{m'n}(\mathbf{k})$, which follows from Eq. (\ref{eq_Zmatrix_definition}).

Here we make a choice of basis for the polarization vectors $\vect{\eta}_\lambda(\mathbf{k})$ -- equivalent to a choice of `electromagnetic' gauge, since the lattice divergence of $A$ is now fixed, using Eq. (\ref{eq_lattice_a}). We take the polarization vectors to be the eigenvectors of $T(\mathbf{k})$ with positive eigenvalues $\zeta_\lambda^2(\mathbf{k})$ (because $T(\mathbf{k})$ is positive-definite for small values of the applied electric field), that is:
\begin{align}
 \sum_m T_{m'm}(\mathbf{k}) \eta_{m \lambda}(\mathbf{k}) = \zeta_\lambda^2(\mathbf{k}) \eta_{m' \lambda}(\mathbf{k}).
 \end{align}
Because of the property $T(\mathbf{-k}) = T(\mathbf{k})^* $, the eigenvectors satisfy $$\eta_{m \lambda} (-\mathbf{k}) = \eta_{m \lambda}^*(\mathbf{k}),$$ and the unitarity of the matrix of eigenvectors implies $$\sum_{m'} \eta_{m' \lambda'}^*(\mathbf{k}) \eta_{m'\lambda}(\mathbf{k})  = \delta_{\lambda \lambda'}.$$

Using these identities, and putting together the electric and magnetic contributions, Eqs. (\ref{eq_app_lattice_e2}) and (\ref{eq_app_lattice_b2}) respectively, we obtain for the full Hamiltonian:
\begin{align}
\mathcal{H}_\text{eff} =& \frac{1}{4} \sum_{\mathbf{k}, \lambda} \bigg[ \left( \frac{\zeta_\lambda^2(\mathbf{k})}{\omega_\lambda(\mathbf{k})}  + \omega_\lambda(\mathbf{k}) \right) a_{\lambda \mathbf{k}}  a_{\lambda \mathbf{k}}^\dagger \nonumber \\
&+ \left(  \frac{\zeta_\lambda^2(\mathbf{k})}{\omega_\lambda(\mathbf{k})}  - \omega_\lambda(\mathbf{k}) \right) a_{\lambda \mathbf{k}}  a_{\lambda, -\mathbf{k}} +  \text{h.c.} \bigg].
\end{align}
For the anomalous terms not conserving photon number to vanish, we enforce:
$$ \omega_\lambda (\mathbf{k}) =  |\zeta_\lambda (\mathbf{k}) |, $$
which is the photon dispersion on the lattice, or Eq. (\ref{eq_lattice_dispersion}) in the main text. The Hamiltonian becomes, as expected,
$$ \mathcal{H}_\text{eff} = \sum_{\mathbf{k},\lambda } \omega_\lambda(\mathbf{k}) \Big[a_{\lambda \mathbf{k}}^\dagger a_{\lambda \mathbf{k}} + \frac{1}{2} \Big]. $$ 

\begin{widetext}
\section{Structure Factors}
\label{appen_structurefactors}

In this section we summarize the calculations leading to Eq. (\ref{eq_structurefactors_0}). To this end, we first express the magnetic field in momentum space as
\begin{align}
b_n (\mathbf{k},t) = \frac{1}{\sqrt{N}} \sum_{\mathbf{r}} e^{-i\mathbf{k} \cdot (\mathbf{r} - \mathbf{t}_n/2)} b_{({\bf r},n)}(t),
\label{eq_bnkt}
\end{align}
where $b_{({\bf r},n)}(t)$ is the magnetic field at the lattice site ${\bf r-t}_n/2$ and at time $t$. As a straightforward generalization of Eq. (\ref{eq_lattice_b}) to arbitrary time, we get

\begin{align}
b_{({\bf r},n)} (t) &= \sqrt{\frac{2}{N}} \sum_{\mathbf{k'}, \lambda, l} \sqrt{\frac{\mathcal{M}_l}{\omega_\lambda(\mathbf{k'})}} \Big[ e^{-i\mathbf{k'} \cdot (\mathbf{r} - \mathbf{t}_n/2)- i \omega_\lambda(\mathbf{k'}) t}  Z_{nl}(\mathbf{k'}) \eta_{l \lambda}(\mathbf{k'}) a_{\lambda \mathbf{k'}}  + e^{i\mathbf{k'} \cdot (\mathbf{r} - \mathbf{t}_n/2) + i \omega_\lambda(\mathbf{k'}) t}  Z_{ln}(\mathbf{k'}) \eta_{l \lambda}^*(\mathbf{k'})  a_{\lambda \mathbf{k'}}^\dagger \Big].
\label{eq_bnrt}
\end{align}
Inserting Eq. (\ref{eq_bnrt}) in Eq. (\ref{eq_bnkt}), we get
\begin{align} 
b_n(\mathbf{k},t) &=  \frac{\sqrt{2}}{4} \sum_{\lambda,l} \sqrt{\frac{\mathcal{M}_l}{\omega_\lambda(\mathbf{k})}} \Big[ e^{-i\omega_\lambda(\mathbf{k}) t} Z_{nl}(-\mathbf{k}) \eta_{l \lambda}(-\mathbf{k}) a_{\lambda, -\mathbf{k}} +  e^{i\omega_\lambda(\mathbf{k}) t}  Z_{ln}(\mathbf{k}) \eta_{l \lambda}^*(\mathbf{k}) a_{\lambda \mathbf{k}}^\dagger \Big],
\end{align}
where we used $ \omega_\lambda(-\mathbf{k}) = \omega_\lambda(\mathbf{k})$, and
$ \sum_{\mathbf{r}} e^{-i (\mathbf{k} \pm \mathbf{k'}) \cdot \mathbf{r}} = \frac{N}{4} \delta_{\mathbf{k},\mp \mathbf{k'}}$.
The two-point correlation function is then given by
\begin{align}
\left< b_m(-\mathbf{k},t) b_n(\mathbf{k},0)\right> &=  \frac{1}{8} \sum_{\lambda, \lambda',l,l'} \sqrt{\frac{\mathcal{M}_l \mathcal{M}_{l'}}{\omega_\lambda(\mathbf{k}) \omega_{\lambda'}(\mathbf{k})}} \left<\left[ e^{-i \omega_\lambda(\mathbf{k}) t} Z_{ml}(\mathbf{k}) \eta_{l \lambda}(\mathbf{k}) a_{\lambda \mathbf{k}} + e^{i \omega_\lambda(\mathbf{k}) t}  Z_{lm}(-\mathbf{k}) \eta_{l \lambda}^*(-\mathbf{k}) a_{\lambda, -\mathbf{k}}^\dagger \right]\right. \nonumber\\
& \hspace{4cm}  \times \left.\left[ Z_{nl'}(-\mathbf{k}) \eta_{l' \lambda'}(-\mathbf{k}) a_{\lambda', -\mathbf{k}} +  Z_{l' n}(\mathbf{k}) \eta_{l' \lambda'}^*(\mathbf{k}) a_{\lambda' \mathbf{k}}^\dagger \right]\right> \nonumber\\ 
&= \frac{1}{8} \sum_{\lambda,l,l'}  \frac{\sqrt{\mathcal{M}_l \mathcal{M}_{l'}}}{\omega_\lambda(\mathbf{k})} \eta_{l' \lambda}^*(\mathbf{k})  Z_{l' n}(\mathbf{k})  Z_{ml}(\mathbf{k}) \eta_{l \lambda}(\mathbf{k})  \times \big[ e^{-i \omega_\lambda(\mathbf{k}) t} \left< a_{\lambda \mathbf{k}} a_{\lambda \mathbf{k}}^\dagger \right> + e^{i \omega_\lambda(\mathbf{k}) t}  \left< a_{\lambda, -\mathbf{k}}^\dagger a_{\lambda, -\mathbf{k}}\right> \big], 
\end{align}
where, in the last equality, we dropped the terms that do not conserve photon numbers and those with cross-polarization, because their expectation value is $0$. We also used the fact that $\eta_{\lambda l}(-\mathbf{k}) = \eta_{\lambda l}^*(\mathbf{k})$ and $Z_{l'n}(-\mathbf{k}) = Z_{nl'}(\mathbf{k})$. 

Since the photons follow Bose-Einstein statistics, in thermal equilibrium at temperature $T$, we have 
\begin{align}
\left< a_{\lambda \mathbf{k}}^\dagger a_{\lambda \mathbf{k}}\right> = n_\lambda(\mathbf{k}) = \frac{1}{e^{\omega_\lambda(\mathbf{k})/T}-1},
\end{align}
and
\begin{align}
\left< b_m(-\mathbf{k},t) b_n(\mathbf{k},0)\right> &= \frac{1}{8} \sum_{\lambda,l,l'}  \frac{\sqrt{\mathcal{M}_l \mathcal{M}_{l'}}}{\omega_\lambda(\mathbf{k})} \eta_{l' \lambda}^*(\mathbf{k})  Z_{l' n}(\mathbf{k})  Z_{ml}(\mathbf{k}) \eta_{l \lambda}(\mathbf{k})  \times \big[ e^{-i \omega_\lambda(\mathbf{k}) t} (1+n_\lambda({\bf k})) + e^{i \omega_\lambda(\mathbf{k}) t}  n_\lambda({\bf k}) \big]. 
\end{align}
Therefore, the dynamic structure factor [Eq. (\ref{eq_dy_sf})] is given by :
\begin{align}
I^{\alpha\beta}({\bf k},\omega)=&\frac{1}{8}\sum_{\lambda,m,n,l,l'}(\hat{\bf t}_m\cdot\hat \alpha)(\hat{\bf t}_n\cdot\hat\beta)\frac{\sqrt{\mathcal{M}_l \mathcal{M}_{l'}}}{\omega_\lambda(\mathbf{k})} \eta_{l' \lambda}^*(\mathbf{k}) Z_{l' n}(\mathbf{k})  Z_{ml}(\mathbf{k}) \eta_{l \lambda}(\mathbf{k}) \nonumber \\
&\times \left[ \delta(\omega+\omega_\lambda(\mathbf{k}))(1+n_\lambda(\mathbf{k})) + \delta(\omega-\omega_\lambda(\mathbf{k}))  n_\lambda(\mathbf{k}) \right],
\end{align}
and the equal-time structure factor [Eq. (\ref{eq_et_sf})] is given by
\begin{align}
	I^{\alpha\beta}({\bf k},t=0)=\frac{1}{8}\sum_{\lambda,m,n,l,l'} \left[ (\hat{\bf t}_m\cdot\hat \alpha)(\hat{\bf t}_n\cdot\hat\beta)\frac{\sqrt{\mathcal{M}_l \mathcal{M}_{l'}}}{\omega_\lambda(\mathbf{k})} \eta_{l' \lambda}^*(\mathbf{k})  Z_{l' n}(\mathbf{k})  Z_{ml}(\mathbf{k}) \eta_{l \lambda}(\mathbf{k}) \right] \coth\left(\frac{\omega_\lambda(\mathbf{k})}{2T} \right).
\end{align}
On taking the zero temperature limit, $\coth\left(\frac{\omega_\lambda(\mathbf{k})}{2T} \right) \rightarrow 1$ for $T \rightarrow 0$, we obtain Eq. (\ref{eq_structurefactors_0}) in the main text.

\vspace{5cm} 

\end{widetext}


\section{Lattice theory near the phase transition}
\label{appen_reg}
\subsection{Calculating the $(\nabla \times b)^2$ term}

In order to take the lattice curl of the magnetic field (living on the direct diamond lattice), we sum over the six bonds of an hexagonal plaquette, taking into account the directedness of the field $b_{(\mathbf{r},n)}$ in a similar way to what was presented in Appendix \ref{appen_lattice_photons}. We get:

\begin{align}
&\left(\nabla_{\hexagon} \times b_{(\mathbf{r},n)}\right)_{(\vect{s},m)} 
=  \sqrt{\frac{2}{N}}  \sum_{\vect{k}, \lambda, l,n}  \sqrt{\frac{\mathcal{M}_l}{\omega_\lambda(\mathbf{k})}} \nonumber \\
& \hspace{0.2cm} \times \Big[   e^{-i\vect{k} \cdot (\vect{s} + \vect{t}_m/2)} Z_{mn}(\mathbf{k})  Z_{nl}(\mathbf{k}) \eta_{l \lambda}(\mathbf{k}) a_{\lambda \vect{k}} + \text{h.c.} \Big],
\label{eq_appen_fone}
\end{align}
and thus, computing the curl of $b$ introduced an extra component of $Z(\mathbf{k})$. We stress that, near the Brillouin zone center, this becomes an extra factor of $\mathbf{k}$ which leads to the quadratic part of the photon dispersion relation.

\subsection{Regularizing the photon frequency}

We now consider the regularized lattice Hamiltonian in Eq. (\ref{eq_lattice_ham_matrixform_expanded_higherorder}), and take $\mathcal{W}_m = \mathcal{W} > 0 $ for simplicity. As discussed in the main text, neglecting the direction dependence of $\mathcal{W}$ does not affect the results below in a qualitative way. The new term becomes, following the calculational steps shown in Appendix \ref{appen_lattice_photons}:
\begin{align}
\mathcal{H}_{(\nabla \times b)^2}
&=  \frac{\mathcal{W}}{4}  \sum_{\vect{k}, \lambda, \lambda', l, l'}  \frac{1}{\sqrt{\omega_\lambda(\mathbf{k})  \omega_{\lambda'}(\mathbf{k})}} \tilde{Z}_{l'l}(\mathbf{k}) \nonumber \\ 
&\times \left[ \eta_{l' \lambda'}^*(\mathbf{k})    \eta_{l \lambda}(\mathbf{k})  \left\{  a_{\lambda \vect{k}} a_{\lambda' \vect{k}}^\dagger + a_{\lambda \vect{k}}  a_{\lambda', -\vect{k}} \right\}  + \text{h.c.} \right],
\end{align}
where we defined the matrix $\tilde{Z}(\mathbf{k})$ with elements given by
$\tilde{Z}_{l'l}(\mathbf{k}) = \sqrt{\mathcal{M}_{l'}} \left[Z(\mathbf{k})\right]^4_{l'l} \sqrt{\mathcal{M}_l}$.

We now add this new term to the lattice Hamiltonian already analyzed in Appendix \ref{appen_lattice_photons}.  Choosing the $\vect{\eta}_\lambda$ to be the eigenvectors of the following Hermitian matrix,
\begin{align}
 \left(T(\mathbf{k}) + \mathcal{W} \tilde{Z}(\mathbf{k}) \right) \vect{\eta}_\lambda(\mathbf{k}) = \zeta_\lambda^2(\mathbf{k}) \vect{\eta}_\lambda(\mathbf{k}),
 \label{eq_appen_fthree}
 \end{align}
with positive eigenvalues $\zeta_\lambda^2(\mathbf{k})$ as before,
and using the unitarity of the $\eta$ matrix, which enforces $\lambda = \lambda'$, we finally have:
\begin{align}
\mathcal{H}_\text{eff} =& \frac{1}{4} \sum_{\mathbf{k}, \lambda} \bigg[ \left( \frac{ \zeta_\lambda^2(\mathbf{k})}{\omega_\lambda(\mathbf{k})}  + \omega_\lambda(\mathbf{k}) \right) a_{\lambda \mathbf{k}}  a_{\lambda \mathbf{k}}^\dagger \nonumber \\
&+ \left(  \frac{ \zeta_\lambda^2(\mathbf{k})}{\omega_\lambda(\mathbf{k})}  - \omega_\lambda(\mathbf{k}) \right) a_{\lambda \mathbf{k}}  a_{\lambda, -\mathbf{k}} +  \text{h.c.}     \bigg],
\end{align}
from which we obtain the dispersion relation of the photons, $\omega_\lambda(\mathbf{k}) = | \zeta_\lambda(\mathbf{k}) |$.

Let us now consider the limit where the speed of light vanishes (that is, when coefficients $\mathcal{M}_m$ go to zero). One has to be careful now because the higher-order operator $(\nabla \times b)^2$ is irrelevant in the RG sense. 
We thus take the limit while keeping 
\begin{align}
\mathcal{W} \sqrt{\mathcal{M}_{l'}} \sqrt{\mathcal{M}_l}
\label{eq_appen_ffive}
\end{align}
 constant. Thus, absorbing these factors into $\mathcal{W}$ \footnote{ Note that the higher-order term included in Eq. (\ref{eq_lattice_ham_matrixform_expanded_higherorder}) is manifestly gauge invariant (hence, so are Eqs. (\ref{eq_appen_fone}-\ref{eq_appen_fthree})). However, the approximation made by taking the renormalized $\mathcal{W}$ (see Eq. (\ref{eq_appen_ffive}) and the discussion below it) as the  \textit{isotropic} regulator leads to Eq. (\ref{eq_iso}), and consequently Eq. (\ref{eq_dispersion_reg_2}) in the main text,  \emph{not} being invariant under the $U(1)$ gauge transformations. This choice of regulator thus generates a \textit{third}, spurious finite-energy mode which has to be discarded, e.g., from the computation of spin structure factors. However, such a spurious mode is not expected to change the reappearance of pinch points when the photon velocity goes to zero, as its dispersion is $\sim k^\gamma$ where $\gamma> 1$. \label{refnote} } leads to the following eigenvalue equation determining the photon frequency:
\begin{align} 
\left(T(\mathbf{k}) + \mathcal{W} \left[Z(\mathbf{k})\right]^4 \right) \vect{\eta}_\lambda(\mathbf{k}) = \zeta_\lambda^2(\mathbf{k}) \vect{\eta}_\lambda(\mathbf{k}),
\label{eq_iso}
\end{align}
where $\mathcal{W}$ is taken to be constant near the phase transition. This is equivalent to Eqs. (\ref{eq_dispersion_reg}) and (\ref{eq_dispersion_reg_2}) in the main text.
\bibliography{biblio}
\end{document}